\documentclass[sigconf]{acmart}

\usepackage{lipsum}  
\usepackage{algorithm}
\usepackage[titletoc,title]{appendix}
\usepackage{enumitem}
\usepackage{caption}
\usepackage{amsmath}
\usepackage{subfigure}
\usepackage{subcaption}
\usepackage{tabularx}
\usepackage{dirtytalk}
\usepackage{listings}
\usepackage{array}
\usepackage{multirow}
\usepackage{graphicx}
\usepackage{booktabs}
\usepackage{makecell}
\usepackage{subcaption}
\usepackage{tabularx}
\usepackage{float}
\usepackage{subfigure}
\usepackage{algpseudocode}
\usepackage{amsmath}
\usepackage{caption}

\newcommand{\rqthreeticketurl}{

\begin{figure*}[t]
    \centering
    \subfigure[Over 54,000 NCIM tickets reported via DMCA]{
        \includegraphics[width=0.48\linewidth, trim=0 0 0 30, clip]{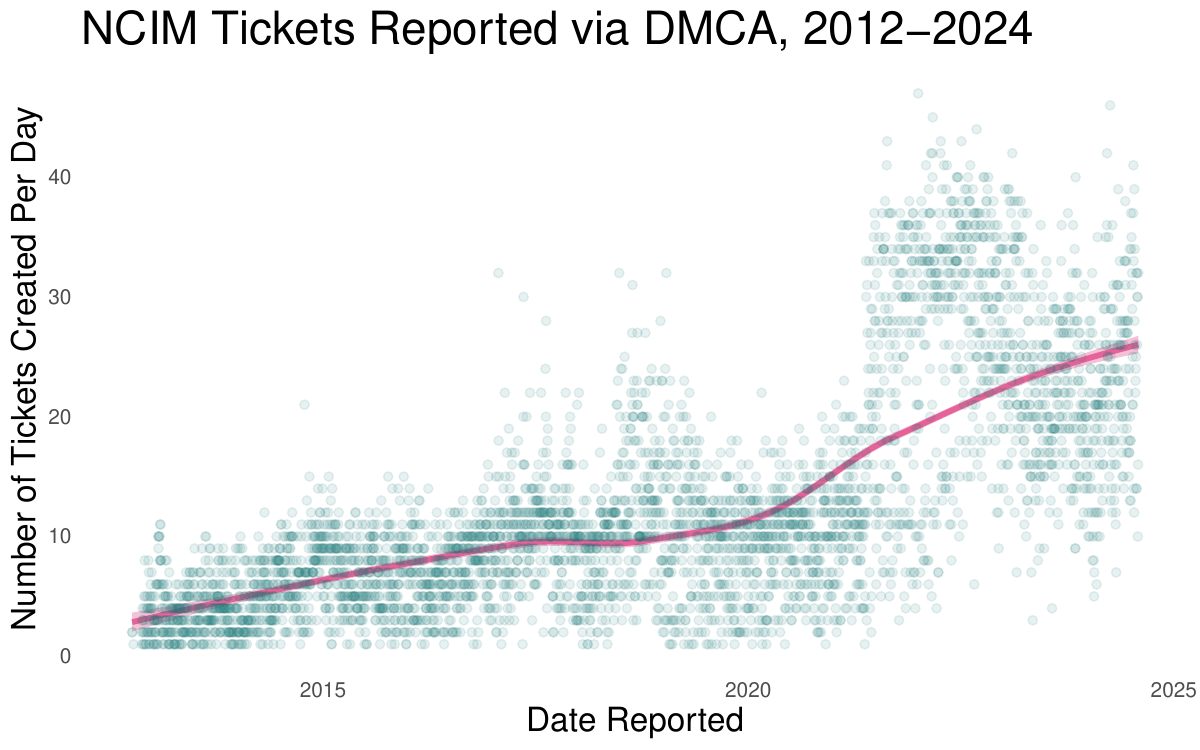}
    }
    \hfill
    \subfigure[Over 85 million NCIM URLs reported via DMCA]{
        \includegraphics[width=0.48\linewidth, trim=0 0 0 30, clip]{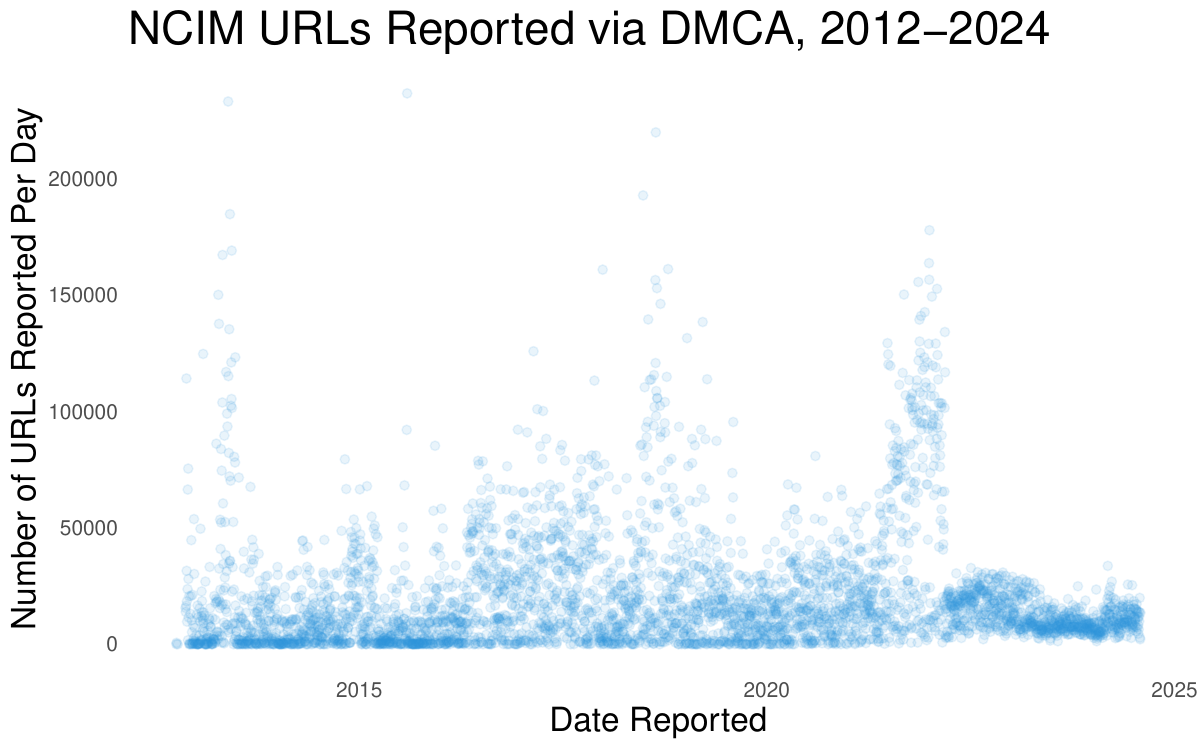}
    }
    \caption{NCIM tickets have increased significantly in last 12 years. URLs have fluctuated with time.}
    \label{fig:rq3-a}
\end{figure*}
}

\newcommand{\rqthreenoncommercial}{
\begin{figure*}[h]
    \centering
    \subfigure[Over 52,000 commercial NCIM tickets reported via DMCA]{
        \includegraphics[width=0.48\linewidth, trim=0 0 0 30, clip]{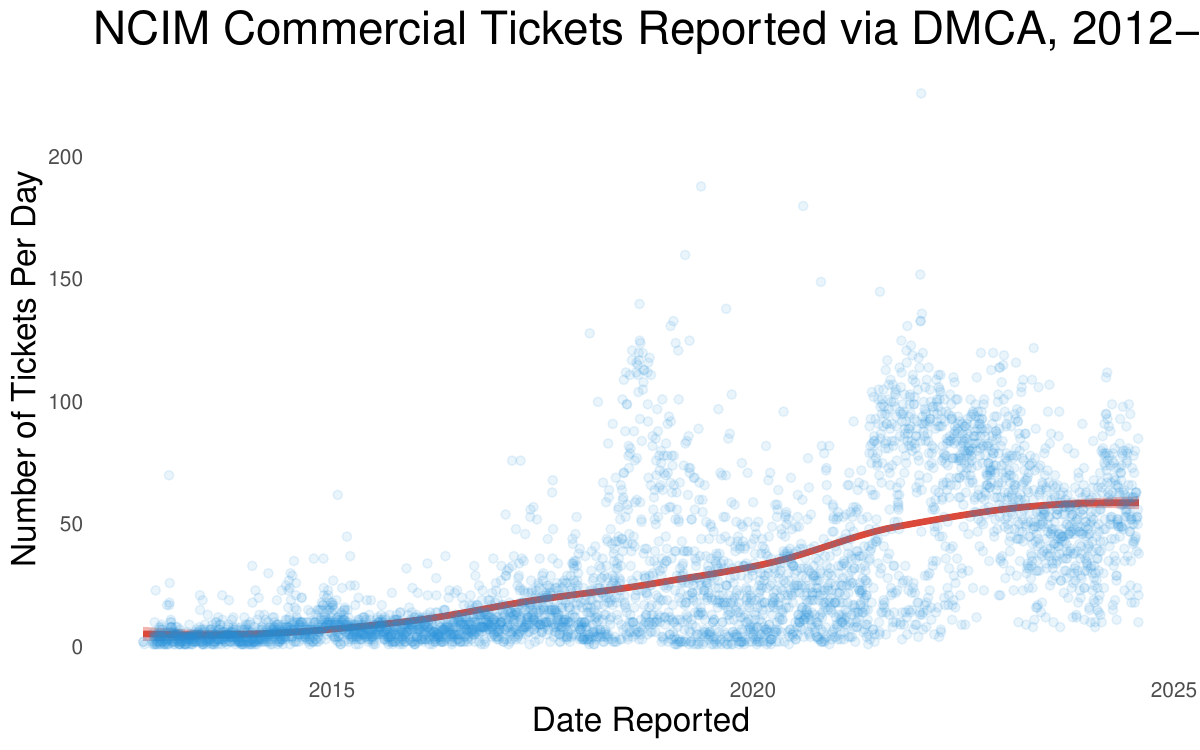}
    }
    \hfill
    \subfigure[Approximately 2,000 non-commercial NCIM tickets reported via DMCA]{
        \includegraphics[width=0.48\linewidth, trim=0 0 0 30, clip]{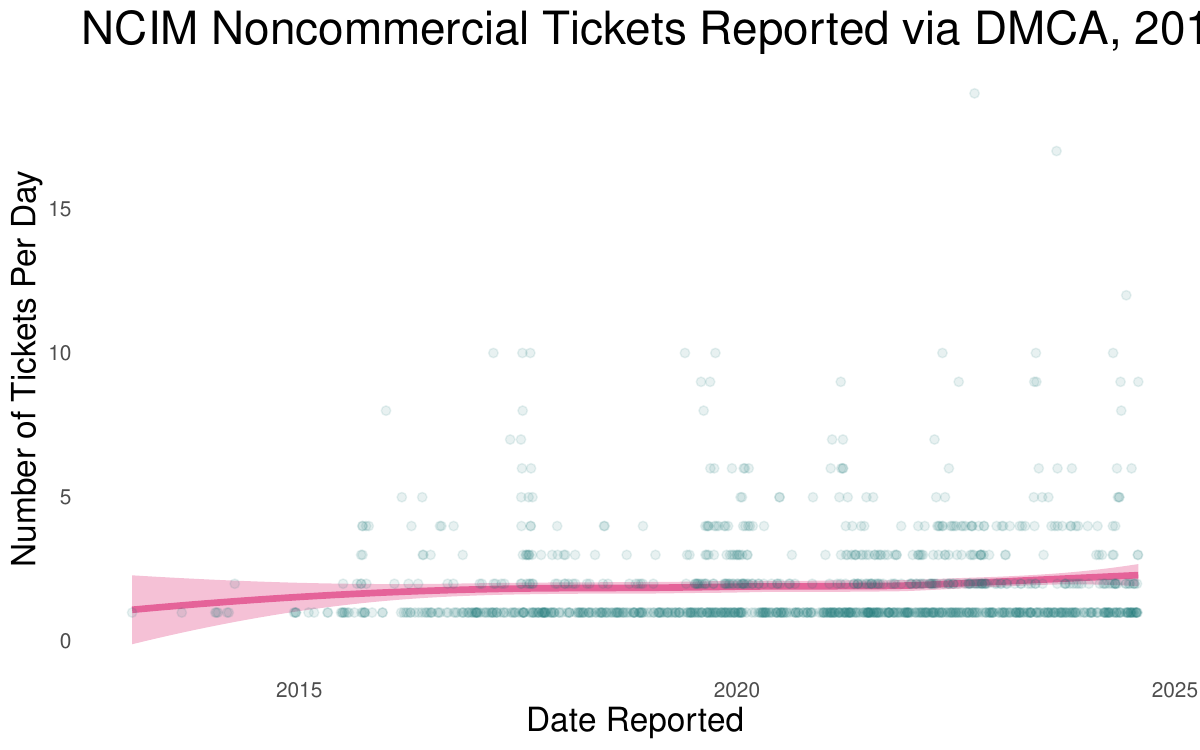}
    }
    \caption{Both commercial and non-commercial NCIM tickets have steadily increased from 2012-2024.}
    \label{fig:rq3-b}
\end{figure*}
}

\newcommand{\keywcommercial}{
    \begin{figure*}[h]
        \centering
            \includegraphics[width=\linewidth, trim=0 0 0 30, clip]{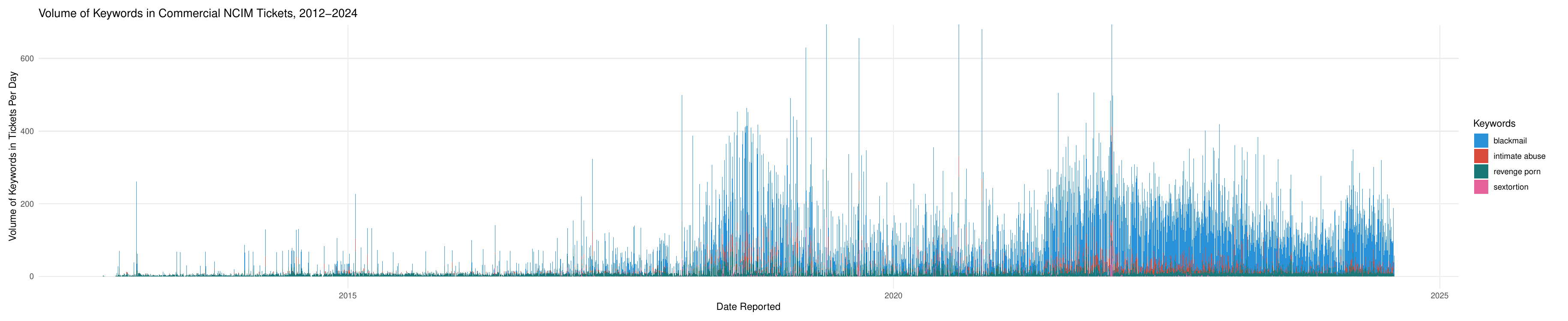}
            \caption{Keywords used to report commercial NCIM using DMCA, 2012-2024 trends}
            \label{fig:rq3-5}
    \end{figure*}
}

\newcommand{\keywnoncommercial}{
    \begin{figure*}[h]
        \centering

            \includegraphics[width=\linewidth, trim=0 0 0 30, clip]{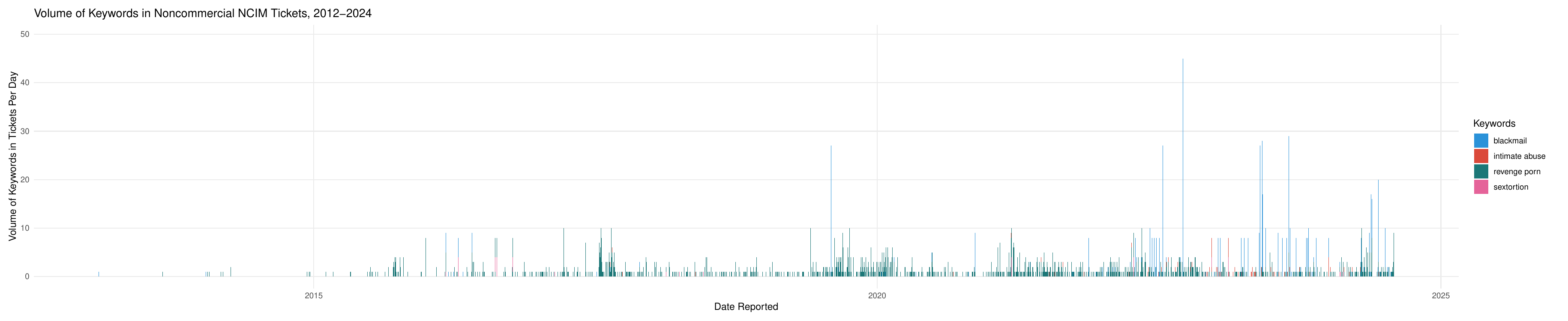}
            \caption{Keywords used to report non-commercial NCIM using DMCA, 2012-2024 trends}
            \label{fig:rq3-6}

    \end{figure*}
}

\newcommand{\lumenUI}{
\begin{figure}[t]
    \centering
    \includegraphics[width=0.5\linewidth, trim=15 220 11 150, clip]{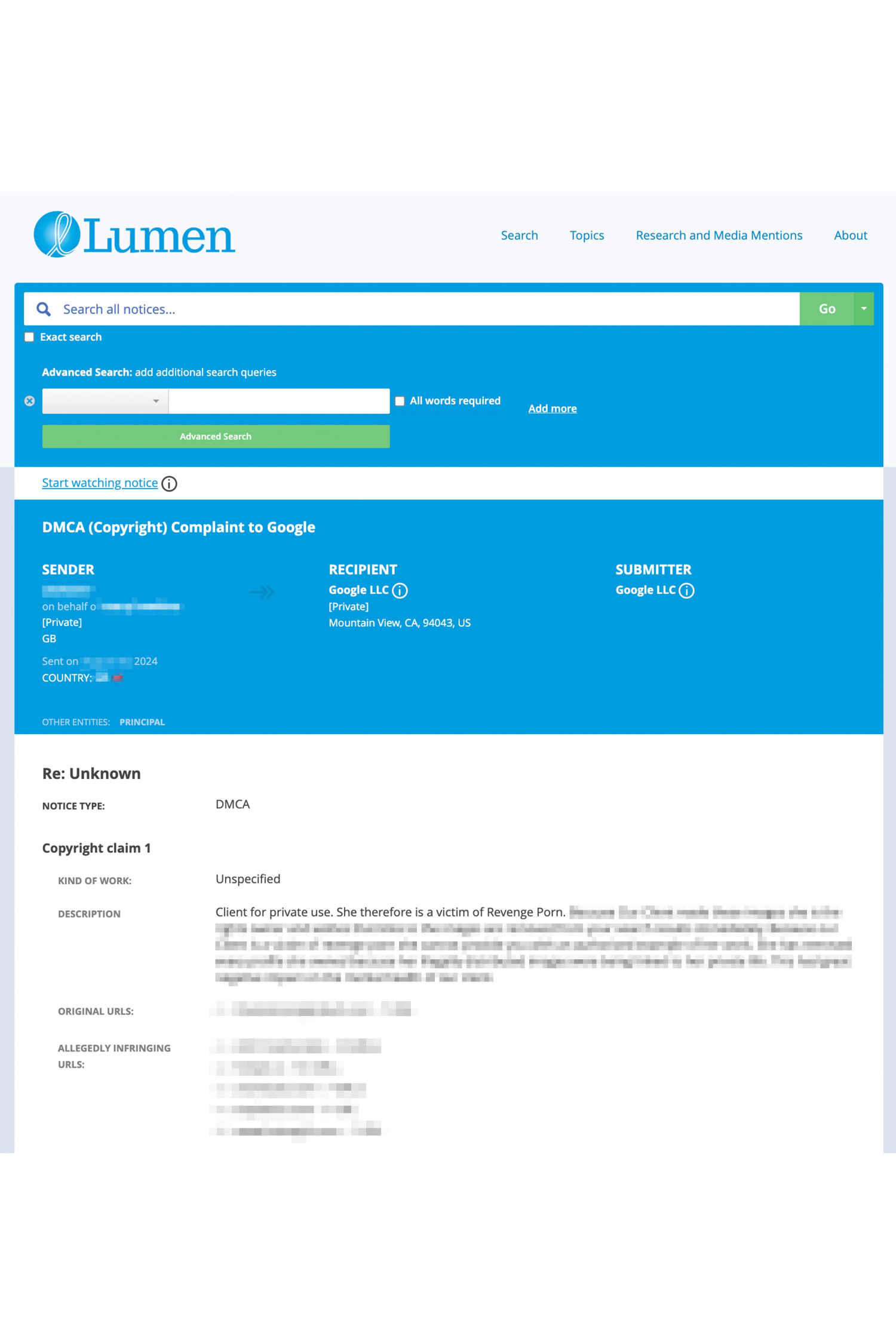}
    \caption{Example of a ticket filed on Lumen database, with identifying information removed. Each ticket contains sender information, which could be a legal or otherwise authorized representative of the copyright holder. The recipient of the ticket is an online platform, such as Google. The platform then submits these tickets to Lumen. Tickets contain a description of the nature of the infringement, which we use the description of the ticket for subsequent qualitative analysis. Tickets include a URL to the original content, and lists URLs which contain the copyrighted content.}
    \label{fig:lumenUI}
\end{figure}
}

\newcommand{\descriptions}{

\begin{table*}[t]
    \centering
    \fontfamily{cmss}\selectfont 
    \begin{tabularx}{0.9\textwidth}{X}

         \textbf{Commercial ticket examples}\\
    \toprule
        \begin{quote}
            \setlength\leftskip{-0.8cm}
            \setlength\rightskip{-0.8cm}
            \textit{``These copyrighted images and videos are materials owned by [redacted], who distributes content solely on official platforms such as OnlyFans. My client strictly forbids distribution outside of official platforms.''}
        \end{quote}
        \vspace{0.05cm}
         \begin{quote}
            \setlength\leftskip{-0.8cm}
            \setlength\rightskip{-0.8cm}
            \textit{``My name is [redacted] and I was a model at [redacted]. A website hosted by your company is now infringing on videos owned by me. This video was posted without my permission.''}
         \end{quote}\\

         \textbf{Noncommercial ticket examples}\\
    \toprule
        \begin{quote}
            \setlength\leftskip{-0.8cm}
            \setlength\rightskip{-0.8cm}
            \textit{``This nude selfie was originally sent to my ex-boyfriend. The platform won't remove it, please remove it from Google Search. It is still cached there and is harming my reputation. This is illegal.''}
        \end{quote}
        \vspace{0.05cm}
        \begin{quote}
            \setlength\leftskip{-0.8cm}
            \setlength\rightskip{-0.8cm}
            \textit{``I am underage in these photos. They are revenge pornography. I do not consent to my full name and images on these websites.''}
        \end{quote}
        \vspace{0.05cm}
        \begin{quote}
            \setlength\leftskip{-0.8cm}
            \setlength\rightskip{-0.8cm}
            \textit{``This user is recording webcams during video chat and using a video of a girl to trick victims to take off their clothes, then blackmailing the victims to pay or threaten to upload their nude videos along with private information, name, addresses, for defamation.''}
        \end{quote}
        \vspace{0.05cm}
        \begin{quote}
            \setlength\leftskip{-0.8cm}
            \setlength\rightskip{-0.8cm}
            \textit{``I took these photos of myself in my underwear, now they are on this website and is damaging me professionally, emotionally, and socially. I am asking for the links to be removed from google images and google search.''}
        \end{quote}\\
    \bottomrule
    \end{tabularx}
    \caption{Examples of paraphrased descriptions for commercial and noncommercial DMCA tickets for NCIM removal.}
\label{tab:dmca_classification}
\end{table*}
}

\newcommand{\algo}{

\begin{figure*}[t]
\centering
\fontfamily{cmss}\selectfont 
\begin{minipage}{0.48\textwidth}
    \centering
    \begin{algorithm}[H]
        \caption{HTTP Status Monitoring}
        \begin{algorithmic}[1]
            \State \textbf{Initial collection}
            \State Collect the last 4 weeks of relevant URLs
            \State Add URLs to table
            \For{each URL in the table}
                \State curl URL for HTTP status code
                \State Write result to the table
            \EndFor
            
            \vspace{3 mm}
            \State \textbf{Collection and monitoring}
            \For{each day}
                \State Collect the current day's URLs
                \State Write URLs to the table
                \For{each URL in the table}
                    \If{last HTTP status == 200}
                        \State Ping the URL for HTTP status code
                        \State Write result to table
                    \EndIf
                \EndFor
            \EndFor
        \end{algorithmic}
    \end{algorithm}
\end{minipage}
\hfill
\begin{minipage}{0.48\textwidth}
    \centering
     \begin{algorithm}[H]
        \caption{Google index monitoring}
        \begin{algorithmic}[1]
            \State \textbf{Initial collection}
            \State Collect the last 4 weeks of relevant URLs
            \State Add URLs to table
            \For{each URL in the table}
                \State Call Custom Search API for URL indexing status
                \State Write result to the table
            \EndFor

            \vspace{3 mm}
            \State \textbf{Collection and monitoring}
            \For{each day}
                \State Collect the current day's URLs
                \State Write URLs to the table
                \For{each URL in the table}
                    \If{last indexing status == True}
                        \State Call API for URL indexing status
                        \State Write result to table
                    \EndIf
                \EndFor
            \EndFor
        \end{algorithmic}
    \end{algorithm}
\end{minipage}
\caption{Processes for monitoring HTTP status and Google Search indexing. Both start by collecting infringing URLs reported in the past 4 weeks, followed by daily updates and monitoring for another 4 weeks. The HTTP monitoring checks if URLs are still active using status codes and keywords to detect removal or redirection. Google Search indexing monitors whether URLs remain indexed using the Google Custom Search API.}
\label{fig:algo}
\end{figure*}
}

\AtBeginDocument{%
  }

\setcopyright{acmlicensed}
\copyrightyear{2025}
\acmYear{2025}
\acmDOI{10.1145/3706598.3713334}
\setcopyright{cc}
\setcctype{by-nc-nd}
\acmConference[CHI '25]{CHI Conference on Human Factors in Computing Systems}{April 26-May 1, 2025}{Yokohama, Japan}
\acmISBN{979-8-4007-1394-1/25/04}

\begin{document}

\title[Inefficacy of the DMCA for Non-Consensual Intimate Media]{A Law of One's Own: The Inefficacy of the DMCA for Non-Consensual Intimate Media}

\author{Li Qiwei}
\email{rrll@umich.edu}
\affiliation{%
  \institution{University of Michigan}
  \city{Ann Arbor}
  \country{USA}
}

\author{Shihui Zhang}
\email{zshihui@umich.edu}
\affiliation{%
  \institution{University of Michigan}
  \city{Ann Arbor}
  \country{USA}
}

\author{Samantha Paige Pratt}
\email{sppratt@umich.edu}
\affiliation{%
  \institution{University of Michigan}
  \city{Ann Arbor}
  \country{USA}
}

\author{Andrew Timothy Kasper}
\email{atkasper@umich.edu}
\affiliation{%
  \institution{University of Michigan}
  \city{Ann Arbor}
  \country{USA}
}

\author{Eric Gilbert}
\email{eegg@umich.edu}
\affiliation{%
  \institution{University of Michigan}
  \city{Ann Arbor}
  \country{USA}
}

\author{Sarita Schoenebeck}
\email{yardi@umich.edu}
\affiliation{%
  \institution{University of Michigan}
  \city{Ann Arbor}
  \country{USA}
}

\renewcommand{\shortauthors}{Qiwei et al.}

\begin{abstract}

Non-consensual intimate media (NCIM) presents internet-scale harm to individuals who are depicted. One of the most powerful tools for requesting its removal is the Digital Millennium Copyright Act (DMCA). However, the DMCA was designed to protect copyright holders rather than to address the problem of NCIM. Using a dataset of more than 54,000 DMCA reports and over 85 million infringing URLs spanning over a decade, this paper evaluates the efficacy of the DMCA for NCIM takedown. \textcolor{black}{Results show that for non-commercial requests, while more than half of URLs are de-indexed from Google Search within 48 hours, the actual removal of content from website hosts is much slower. The median infringing URL takes more than 45 days to be removed from website hosts, and only 5.39\% URLs are removed within the first 48 hours.} Additionally, the most frequently reported domains for non-commercial NCIM are smaller websites, not large platforms. We stress the need for new laws that ensure a shorter time to takedown that are enforceable across big and small platforms alike.

\end{abstract}



\keywords{Online sexual abuse, Content moderation, Image-based sexual abuse, IBSA, Non-consensual intimate imagery, NCII, DMCA, Digital copyright, Survival analysis}

\received{12 September 2024}
\received[revised]{10 December 2024}
\received[accepted]{16 January 2025}

\maketitle

\section{Introduction}

Non-consensual intimate media (NCIM) refers to the unauthorized creation, obtainment, or distribution of sexual content featuring someone's body or likeness~\cite{qiwei2024sociotechnical}. NCIM can take various forms, including unauthorized dissemination of commercial content, sexualized deepfakes, and what is colloquially known as ``revenge porn''. NCIM predominantly affects women, who account for approximately 90\% of all victim-survivors~\cite{eaton20172017,ccri2014revenge,safedigitalintimacy}. The psychological impact of NCIM is severe, with nearly all victim-survivors experiencing severe emotional and social distress~\cite{ccri2014revenge}. Around half contemplate suicide, and some tragically take their own lives to escape the emotional pain and social stigma~\cite{compton2024deepfake,ccri2014revenge,eaton20172017}. 

While there is broad recognition of NCIM harms and some consensus on the need for regulation, significant uncertainties remain on \textit{how} to address NCIM~\cite{citron2014criminalizing,de_angeli_reporting_2023,citron2020internet,franks_drafting_2014}. In particular, deepfake videos---98\% of which are pornographic---are often weaponized to silence individuals---primarily women, including public figures like reporters and lawmakers~\cite{securityhero_state_of_deepfakes_2024,compton2024deepfake}. This gendered harm also represents a broader societal threat, as anyone can be targeted. Nearly every state in the U.S. has enacted laws targeting NCIM. While these laws are a critical step forward, they primarily focus on prosecuting the initial perpetrator, falling short in addressing the pressing concern for NCIM victim-survivors: the removal of harmful content from the internet. Further compounding the issue is Section 230 of the Communications Decency Act (CDA), which shields online platforms from liability for third-party content, effectively absolving them of responsibility for hosting NCIM. Given the ease of distribution, lack of barriers to duplication, and the amplification from social computing elements---such as content algorithms, messaging platforms, social media networks---NCIM continues to spread rapidly online~\cite{qiwei2024sociotechnical}. As a result, despite NCIM being recognized as a crime in most states, victim-survivors find themselves lacking effective tools to remove abusive content from websites.

One of the few tools available for NCIM removal is the Digital Millennium Copyright Act (DMCA). Enacted in 1998, the DMCA mandates that U.S.-based websites respect digital copyright, protecting creative works by allowing copyright holders to request the removal of infringing content~\cite{dmca_1998}. It is widely used by publishers, media companies, and artists, resulting in the removal of large amounts of content online~\cite{garcia_v_google_2015,seng2021copyrighting,fromer2015should}. As of January 2024, Google received an estimated 30 million DMCA requests per week~\cite{gigazine_google_dmca_2024}. \textcolor{black}{It's important to note the DMCA has faced widespread criticism for stifling free speech and enabling powerful media companies to exert control over internet content~\cite{salam2012copyright,carpou2015robots,hollister2024nintendo}. Despite this complicated backdrop, some }scholars have looked towards DMCA as a tool for NCIM removal, arguing that victim-survivors can assert copyright ownership over their content to protect sexual privacy. Through this mechanism, the DMCA has been used to remove NCIM~\cite{farries2019feminist,ccri_online_removal}. However, this approach reveals a critical misalignment: The DMCA is designed to protect economic interests, while NCIM victim-survivors are primarily concerned with preserving their privacy and psychological well-being. Although the DMCA has been effective in some cases in removing NCIM, its application is flawed and can even exacerbate challenges. Filing a DMCA report requires disclosing personal information, which is shared with the infringing party, further compromising privacy. Moreover, copyright is typically granted to the person who created the recording, which in some cases may be the abuser~\cite{d2015fighting,farries2019feminist,franks_drafting_2014}. Finally, even when content is removed, it is often re-uploaded repeatedly~\cite{franks_reforming_2021}.

These issues highlight a broader problem with NCIM reporting, which, despite being part of content moderation, has received little research attention~\cite{qiwei2024sociotechnical}. Few studies examine how NCIM is reported or how quickly it is removed, especially through legal mechanisms like the DMCA. This is concerning given the scale of NCIM, in-part driven by rapid increases in deepfakes~\cite{securityhero_state_of_deepfakes_2024,compton2024deepfake}. Particularly, laws around AI copyright and deepfake removals online are nascent or nonexistent, making deepfake NCIM a particularly urgent and unaddressed issue~\cite{compton2024deepfake,lee2023talkin,oversightboard_deepfake_2024}.

We present the first in-depth analysis of DMCA takedown requests related to NCIM. Our study includes a survival analysis of URLs to measure the time for NCIM to be removed from web hosts, and deindexed from Google Search---crucial steps in mitigating harm. The harm caused by NCIM can escalate rapidly, making swift removal crucial. A common benchmark for reducing exposure and limiting damage is to remove the content within 48 hours~\cite{oversightboard_deepfake_2024,take_it_down_2024}. Throughout this paper, we adopt this 48-hour benchmark to assess the effectiveness of content removal.
We use \textit{Time-To-Takedown (TTT)}---time between when content is reported and when it is removed or deindexed, measured in days---to assess the efficacy of the DMCA. Additionally, we conduct a longitudinal analysis of NCIM reports from 2012 to 2024 and identify characteristics of infringing platforms and hosts. This study is guided by the following research questions:

\begin{list}{}{\leftmargin=1cm}
    \item \textbf{RQ 1: Web hosts TTT} How effective is the DMCA for removing NCIM from websites?
    \item \textbf{RQ 2: Google Search TTT} How effective is the DMCA for deindexing NCIM from Google Search?
    \item \textbf{RQ 3: Analysis of tickets 2012-2024} How can we characterize the growth of NCIM and infringing platforms? 
\end{list}

We collected a large dataset of DMCA takedown requests from Lumen, a database hosted by Harvard's Berkman Klein Center.\footnote{https://lumendatabase.org/} In the first part of the study, we analyzed 1,564 tickets involving 70,502 infringing URLs, tracking their TTT from web hosts and Google Search for over eight weeks. The majority of tickets surfaced are \textit{commercial}---filed by performers for economic reasons, and a smaller portion are \textit{non-commercial}---including what is commonly referred to as ``revenge porn''. We show that \textcolor{black}{about 40\% of URLs remain accessible on web hosts 60 days after being reported, with only 5.39\% }removed within the key 48 hours after reporting. Google Search deindexing occurs at a faster pace, with \textcolor{black}{52.27}\% of URLs deindexed 48 hours after reporting. In the second part of the study, we examined NCIM-related DMCA reports from 2012 to 2024, covering more than 85 million URLs from over 54,000 tickets. Results show that smaller web hosts are responsible for most infringements, highlighting the need for moderation efforts beyond the large major platforms.

Our findings have significant implications for NCIM removal policies. Legal scholars emphasize the urgent need for NCIM-specific legislation~\cite{citron2014criminalizing,citron2020internet,citron2022fight,franks_reforming_2021}. Policies beyond copyright law are necessary to hold web hosts accountable for the harms caused by NCIM. New laws must balance the urgent need for rapid content removal for victim-survivors with the legal requirements of due process. Additionally, they must address the DMCA’s shortcomings in managing emerging threats like deepfakes and provide the flexibility to override Section 230 when appropriate.

\section{Background}
To understand the challenges of addressing NCIM, we first consider the legal frameworks that have been adapted to tackle this issue. Although NCIM is recognized for its severe impact on victim-survivors' privacy and well-being, the legal response for removal has largely been through the lens of intellectual property (IP) law~\cite{gilden2019copyright}. Central to this is the Digital Millennium Copyright Act (DMCA), a key piece of legislation within IP law. Originally intended to protect creative works, the DMCA has increasingly been used---though imperfectly and controversially---to facilitate the removal of NCIM from online platforms~\cite{gilden2019copyright,gilden2018sex,fromer2015should}. Three predominant concerns emerge at the intersection of NCIM and IP law. First, current legal methods for NCIM removal often fail to address many kinds of NCIM, specifically non-commercial NCIM (what is known as ``revenge porn''). Second, IP ``ought'' to be reserved for the protection of \textit{commercial} interests. Finally, because IP law has been somewhat effective in removing NCIM, lawmakers and technologies alike have failed to develop more robust processes to address NCIM.  

\subsection{Non-consensual intimate media}
As digital networks and recording technologies become more integrated into people's sexual lives, they introduce a range of new risks for users~\cite{gilden2018sex,ccri_online_removal,citron2018sexual,citron2014criminalizing,geeng_usable_nodate,tseng2020tools,tseng2022care}. Non-consensual intimate media (NCIM) is a form of online sexual abuse that commonly targets women but can happen to anyone. While ``revenge porn'' is the most well-known form of NCIM, the issue is broad, encompassing sexualized deepfakes, non-consensual filming, and distribution of explicit content~\cite{henry2024image,eaton20172017, eaton2024victim,flynn2019image}. The core issue revolves around \textit{consent}. When a perpetrator removes intimate content from the context in which it was originally shared---such as a private agreement between two people---this act constitutes a violation~\cite{rosenberg2022revenge}. For example, if someone says, ``This video is only for you and me during our romantic relationship'', and the content is shared beyond that context, it breaches that agreement~\cite{nissenbaum2004privacy,citron_criminalizing_2016}. The consequences of NCIM are severe and multifaceted, often leading to significant psychological distress such as anxiety and depression, and in extreme cases, driving victims to end their lives~\cite{batool_expanding_2024, ccri2014revenge, mcglynn_its_2021,eaton2024victim,srinivasan2021right}.

There are also some common ``myths'' about NCIM. Two of the most common misconceptions are the belief that sharing NCIM is a legitimate exercise of free speech, and the assumption that deepfake NCIM causes only ``personal'' harm as opposed to a ``societal issue'' or ``threat to democracy'', making them less significant. First, NCIM and ``revenge porn'' are not protected by the First Amendment as they do not involve public discourse~\cite{koppelman2015revenge,rosenberg2022revenge,us_copyright_office_digital_replicas_2024}. Various statutes provide exceptions for conduct that might involve protected speech, such as news reporting, political campaigns, commentary, and satire~\cite{us_copyright_office_digital_replicas_2024}---NCIM does not fall within these protected categories. Second, the rise of sexualized deepfakes represents a significant societal threat, particularly to democracy. Nearly all (98\%) deepfakes online are pornographic and are often weaponized to target and silence individuals, including political reporters, lawmakers, and other prominent public figures~\cite{securityhero_state_of_deepfakes_2024,schumer_roadmap_nodate, nist_synthetic_content_2024}. The ability to target any individual---while often a gendered harm disproportionately affecting women---can impact anyone, making this a widespread societal issue. It threatens the freedoms of those who are targeted, as well as those who may be threatened with becoming future targets~\cite{freed2018stalker,mcdonald_its_nodate,brigham2024violation,jacobsen_deepfakes_2024}.

The harms of NCIM highlight the need for rapid removal. The surge in attention during the early hours after content is posted translates directly into harm~\cite{mcglynn_its_2021,eaton2024victim}. The Oversight Board, which oversees content moderation at Meta, declared that non-consensual deepfakes must be removed within 48 hours~\cite{oversightboard_deepfake_2024}. Similarly, the TAKE IT DOWN Act mandates platforms address reported content within 48 hours~\cite{take_it_down_2024}. Research shows that online information often reaches saturation within 30 hours~\cite{salvania2015information}, and popular hashtags on platforms like Twitter (now X) lose attention after just 17 hours~\cite{lorenz2019accelerating}. Considering these various timelines, we take the most conservative one at 48 hours. In the remainder of the paper, we use this 48-hour window as a benchmark for assessing removal speeds. 

As of 2024, 48 states of the U.S. and Washington D.C. have some form of law targeting the initial perpetrator. In 2022, Congress also passed civil legislation targeting the disclosure of intimate media as part of the Consolidated Appropriations Act~\cite{US_RepublicAct_HB2471_2022}. Victim-survivors can bring forth lawsuits when their content is distributed by another party online. Similarly, the legal landscape for sexualized and pornographic deepfakes is developing. Ten states so far have legislation to target perpetrators who create and share deepfakes~\cite{Texas_SB1361_2023, SD_Statute_22_21_4, NY_SB1042_2023}. 

While these laws represent significant progress, they are often too narrow and lack sufficient enforcement power to address the full scope of the harm. Even though civil suits allow victim-survivors to seek compensation when their content is shared by downstream distributors, identification and prosecution can be challenging. Platforms must cooperate by providing the distributors' user logs, and their IP addresses must be traceable. Furthermore, much of the harm occurs in private channels, such as text messages or end-to-end encrypted chats on platforms like Discord, Telegram, WhatsApp, and Signal~\cite{mcglynn_its_2021,scheffler2023sok}. This makes locating and surfacing the content particularly difficult. Deepfake laws present similar enforcement challenges. Many state laws require that the intent to cause harm be proven, which can be difficult if the perpetrator claims the deepfakes were created ``for entertainment''~\cite{USNews_DeepfakeBan_2024,us_copyright_office_digital_replicas_2024}. 

In essence, although legislation exists, legal action currently cannot ensure that infringing content will be removed from the internet. This is the central pain point of victim-survivors, who fear the reappearance of content online, causing repeated disruption in their lives, including doxxing, harassment, and other kinds of abuses~\cite{mcglynn_its_2021,ccri2014revenge,citron2020internet,eaton2024victim}. \textcolor{black}{Section 230's broad exemptions for online platforms have led to severe consequences across multiple domains. Not only does it shield websites that host abusive content from legal accountability, but it has also enabled platforms like Facebook to evade responsibility even in cases of facilitating mass atrocities, as evidenced by its role in the genocide of the Rohingya community in Myanmar~\cite{sloss2020section,nourooz2023transitional,johnson2018beyond}. In the context of NCIM specifically, victim-survivors are unable to take legal action against the platforms that continue to host their infringing content~\cite{uscode_section_230_2012, cda_section_230, gilden2018sex}.} To this end, \textcolor{black}{some }scholars have argued that Section 230 is an impediment to justice and should be revised to address these harms~\cite{citron2023fix, citron2020internet}. \textcolor{black}{Yet at the same time, the American Civil Liberties Union argues that Section 230 helps protect online users First Amendment rights, noting that a world where online platforms were accountable for user-generated content would end online exchanges and discussions~\cite{aclu_section230_2020}. These challenges and complexities }underscore the need for more effective and nuanced approaches to combating NCIM \textcolor{black}{using the law}.

\subsection{Intellectual property}
Copyright law in the United States is designed to protect and promote creative labor~\cite{feist_v_rural_1991,garcia_v_google_2015,gilden2019copyright,fromer2015should}. It ensures that one party's work cannot be easily appropriated without due process or compensation~\cite{gilden2018sex}. With the advent of the internet and easy content duplication and distribution, the Digital Millennium Copyright Act (DMCA) was introduced to extend IP protections into the digital domain. Under the DMCA, copyright holders make reports about infringing content, and U.S.-based website operators are legally obligated to respond to legitimate requests by removing infringing content~\cite{dmca_1998,seng2021copyrighting}.

\color{black}
However, many scholars argue that the DMCA has evolved into a mechanism for control and censorship online~\cite{cobia2009digital,salam2012copyright}. The law grants copyright holders significant authority over online content, often with implications that extend beyond traditional IP concerns~\cite{cobia2009digital}. The DMCA creates a complex web of power among diverse stakeholders---large corporations, independent creators, platform operators, and internet users---whose interests often conflict. Below, we summarize some key criticisms of the DMCA. 

\subsubsection*{Volume and legitimacy. }Much of the wider discourse surrounding the DMCA focuses on the (il)legitimacy and high volume of automated DMCA requests. These ``robo-takedowns'' can stifle user-generated content, even when users have little intention (or awareness) of their infringement~\cite{fiesler_lawful_2020}. Illegitimate takedowns notices can even sometimes be initiated by parties who don't actually own the copyright~\cite{cobia2009digital}. While the DMCA does provide a counter notice mechanism allowing recipients to justify their usage, Carpou notes the counter notice procedure is rarely utilized in reality, leaving users with limited ability to respond to takedown notices~\cite{carpou2015robots}.

% 1)  censorship to free speech 
\subsubsection*{Censorship. }Another central concern is the DMCA's potential to enable censorship, undermine free speech, and allow corporate overreach that extends far beyond the law's intended scope~\cite{radsch2023weaponizing,hollister2024nintendo,fiesler_chilling_2023}. The DMCA includes fair use provisions that carve out exemptions for using copyrighted content for purposes such as commentary, news reporting, and teaching~\cite{dmca_1998}. However, how these protections are implemented may be problematic in practice. Submitting a DMCA removal request requires no judicial oversight---the internet host serves as the arbiter. However, the same online host would face legal liability if it fail to remove infringing content. Combined, this creates a strong incentive for online hosts to err towards removals~\cite{cobia2009digital}. 

% unfair protection for larger companies 
\subsubsection*{Unequal protection. }The DMCA's enforcement mechanisms disproportionately benefit large copyright holders---typically powerful media companies---over smaller entities or independent creators. While companies like Disney, Sony, or Nintendo can swiftly issue DMCA takedowns through automated systems and teams of lawyers~\cite{hollister2024nintendo}, independent copyright holders often lack the resources to protect their IP effectively. Consequently, the DMCA system primarily serves the interests of the established media industry, creating an uneven playing field that reinforces existing power dynamics in content creation and distribution~\cite{cobia2009digital}.

% hinderance to usage of technology 
\subsubsection*{Hindrance to technology use. }DMCA has been noted to induce ``chilling effects'' on online creators---which in this case, creators refers to those who create literature or art, not NCIM~\cite{fiesler_chilling_2023}. Fiesler argues that users see the law as a barrier to their desired use of technology, and at the same time the website meant to facilitate creativity and sharing do not properly support users in navigating copyright~\cite{fiesler2015understanding,fiesler_lawful_2020}. Additionally, the DMCA has prevented security researchers from investigating bugs in software~\cite{sardaryzadeh2022security}. Overall, many scholars in tech policy believe the DMCA allows a ``free pass'' for corporate overreach on the internet~\cite{sardaryzadeh2022security,radsch2023weaponizing}, attribute issues to both the fair use doctrine and under-utilization of the counter notice~\cite{cobia2009digital}. 

% transition sentence
While these criticisms highlight the DMCA's shortcomings, the law has also evolved to serve purposes far beyond its original scope. Gilden highlights ``a growing contradiction at the core of copyright law'', where despite courts and scholars often claiming that copyright is primarily concerned with authors' economic interests, it increasingly serves to protect ``interests such as privacy, sexual autonomy, reputation, and psychological well-being \dots it just uses the language of money and markets to do so''~\cite{gilden2019copyright}. While courts have historically resisted extending these protections to personal and non-commercial harms, the rise of social media has amplified the relevance of intellectual property law in these contexts~\cite{gilden2018sex}. As Gilden points out, the threshold for copyright protection is low, requiring only a ``modicum of creativity''. As a result, everyday content---like Facebook posts, Instagram photos, or YouTube videos created on smartphones---can give rise to new copyright interests~\cite{feist_v_rural_1991, gilden2018sex}. In return, copyright law has also offered individuals the opportunity to push back against the platforms that profit from their vulnerabilities~\cite{cohen_platform_economy_2017, gilden2018sex}.

\color{black}

\subsection{Using intellectual property law for NCIM}

Victim-survivors leverage the protections granted to IP owners to have non-consensual content removed from the internet~\cite{d2015fighting, farries2019feminist}. Gilden notes ``The powers created by IP laws---namely the ability to substantially control the use of names, images, voices, and texts---implicate far more than intellectual production''~\cite{gilden2018sex}. The same laws put forth to protect financial interests are also a useful ``toolkit'' to claim property rights online~\cite{gilden2018sex}. Notably, Section 230 does \textit{not} shield platforms against IP infringements: even though Section 230 protects platforms from content posted by their users, platforms still must respond to requests of copyright holders~\cite{gilden_copyrights_nodate,gilden2018sex}. In other words, protections offered by the DMCA supersede that of Section 230.

\begin{figure*}[t]
    \centering
    \includegraphics[width=0.85\linewidth, trim={10 50 10 20},clip]{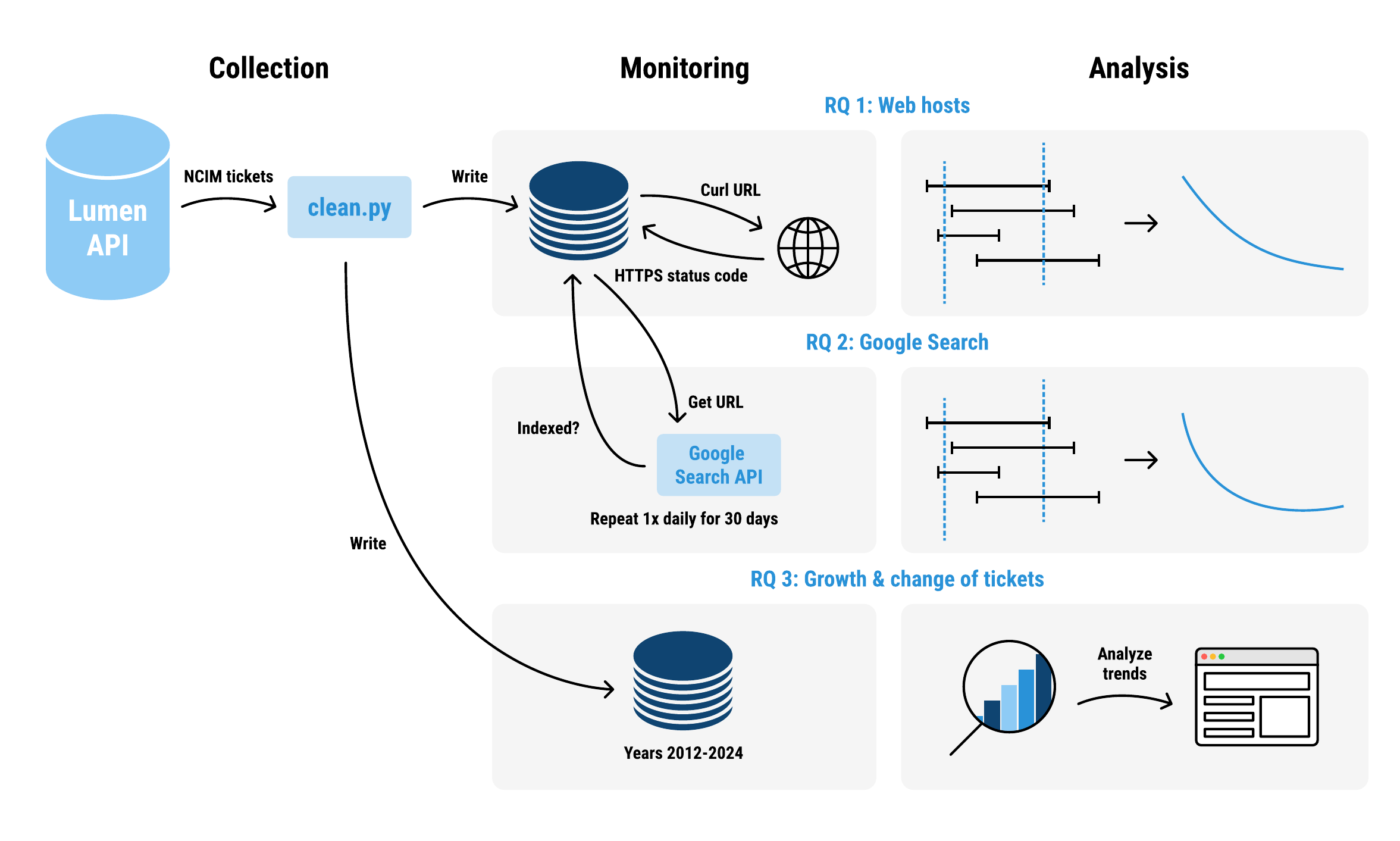}
    \caption{\textit{Method---}Data collection, monitoring, and analysis. We query Lumen API for relevant NCIM tickets, which are then processed by a Python script that extracts infringing URLs and monitors them over a span of 4 weeks. The URLs are tracked for HTTP status via curl to the web page and Google Search indexing via the Google Search API. We use this data to answer: \textit{RQ 1:} time-to-takedown (TTT) from web hosts \textit{RQ 2:} TTT from Google Search \textit{RQ 3:} changes and characteristics of NCIM reports and web hosts, 2012-2024. }
    \label{fig:lumen-hero}
\end{figure*}

Due to concerns of ``chilling effects'' and IP law dilution, copyright scholars are typically reluctant to expand the legal definition of IP laws to cover NCIM and other personal usages~\cite{fromer2015should,gilden2019copyright,fiesler_chilling_2023}. This results in spotty legal coverage for removals that do not have a financial implication for the copyright holder~\cite{us_copyright_office_digital_replicas_2024}. \textcolor{black}{ However, Smith argues, copyright law has become the default tool for preventing the dissemination of intimate content precisely because of limitations in other legal frameworks. While privacy laws exist, they often fail to provide the immediate relief that NCIM victim-survivors desperately need, making DMCA notices the most practical option available~\cite{smith2021weaponizing,ccri_online_removal}}. Thus, using the DMCA for NCIM removal remains somewhat controversial, but lacking a better alternative, DMCA has been a valuable tool for victim-survivors~\cite{gilden2018sex, franks_reforming_2021}. 

Even in the best-case scenario, there can be drawbacks to using the DMCA for NCIM takedown. First, a DMCA report requires the victim-survivor to divulge personal information, including legal name, address, and contact information. This information is sent to the user who posted the content, who in many cases are abusers. Second, individuals may be hesitant to take this step because of the emotional shame and public shaming involved---``revenge porn websites are specifically aimed at shaming individuals and ruining their reputations. Issuing a takedown notice may have the opposite effect: it can draw more attention to these images and encourage their re-uploading to alternative sites, necessitating multiple takedown notices and potentially leading to further online abuse such as doxing''~\cite{o2020using}. Third, not all NCIM is created by the depicted person, meaning that victim-survivors do not always own the copyright to the images improperly posted online~\cite{d2015fighting}. Fourth, many web platforms are hosted outside the United States, rendering the threat of copyright infringement ineffective and enforcement practically impossible~\cite{d2015fighting}. Finally, DMCA requests will only remove the specific instance of the content. Even when a site owner complies with a takedown request, the same content is often reposted on another platform or reappears on the same site. This forces victim-survivors to repeatedly go through the reporting process~\cite{franks_reforming_2021,franks_sex_nodate}.

\section{Method}

This study examines the effectiveness of the DMCA for removing NCIM from the internet. Our approach involves identifying and categorizing NCIM-related tickets, monitoring the status of URLs to determine how long infringing content remains accessible online, and tracking the time it takes for Google Search to deindex these URLs. We perform a survival analysis to estimate the speed of content removal. Finally, we examine trends in NCIM reporting over a 12-year period to understand the evolving landscape of NCIM harms.

\subsection{Research questions}

We use a 48-hour benchmark based on prior research to determine the efficacy of the DMCA for addressing NCIM~\cite{oversightboard_deepfake_2024,lorenz2019accelerating}. Our research questions compare URL time-to-takedown (TTT) with the 48-hour guideline. First, we measure the TTT of infringing URLs upon reported to web hosts. At the same time, we measure TTT for Google Search. Search engines can deliver NCIM harms by making abusive content searchable through victim-survivors' names or reverse-image search, making fast action crucial~\cite{bates_revenge_2017,mcglynn_its_2021}. For both sets of data, we examine differences between commercial and non-commercial content. Commercial content, typically submitted by models and sex workers, leverages the DMCA as intended---to protect works and financial interests of copyright holders. Non-commercial tickets, although not designed for DMCA use, highlight the tool’s importance for removing abusive content. Finally, we conduct a longitudinal analysis (2012-2024) of tickets, URLs, and infringing domains. We track the evolution of DMCA reports for NCIM and identify key characteristics of the worst infringing websites. Our research questions are: 

\begin{enumerate}
    \item \textit{Web hosts TTT: How effective is the DMCA for removing NCIM from websites?} 
    \begin{enumerate}
        \item[1.1] How many days does it take to remove NCIM from web hosts using the DMCA?
        \item[1.2] What proportion of content is removed within the first 48 hours after reporting?
        \item[1.3] Do removal timelines differ between commercial and non-commercial content? 
    \end{enumerate}

    \item \textit{Google Search TTT: How effective is the DMCA for deindexing NCIM from Google Search?}
    \begin{enumerate}
        \item[2.1] How many days does it take to deindex NCIM from Google Search using the DMCA?
        \item[2.2] What proportion of content is deindexed within the first 48 hours after reporting?
        \item[2.3] Do removal timelines differ between commercial and non-commercial content?
    \end{enumerate}

    \item \textit{Analysis of tickets 2012-2024: How can we characterize the growth of NCIM?}
    \begin{enumerate}
        \item[3.1] How has the volume of reports changed in the last 12 years?
        \item[3.2] How many websites host non-commercial NCIM?
        \item[3.3] What are some characteristics of these websites?
    \end{enumerate}
\end{enumerate}

\subsection{Sending and collecting DMCA tickets}

We collected data from Lumen, a database that archives millions of DMCA takedown requests. Lumen was founded in 2002 by legal scholar Wendy Seltzer, originally to track copyright claims, but then expanding to document and analyze takedown notices related to online content broadly~\cite{seltzer2010free}. Its primary purpose is to understand the various types of complaints that platforms receive. It is now managed by the Berkman Klein Center for Internet and Society at Harvard University. Although Lumen hosts numerous NCIM-related requests, it does not conduct specific research on NCIM and does not provide tagging or categorization methods to distinguish NCIM from other types of copyright requests. These requests, labeled as ``tickets'', are submitted by individuals and other entities seeking content removal. Once a ticket is created and sent to a platform---such as Google, Wikipedia, or Vimeo---the platform forwards it to Lumen for archiving. It is important to note that a ticket's presence in Lumen does not indicate whether the removal request was granted or denied by the platform; Lumen merely records the requests without documenting or influencing the outcome. Each DMCA request in the Lumen database is cataloged as a unique ticket with a distinct ID. A single ticket may include one or more infringing URLs, which point to the specific content the reporter seeks to remove. Typically, tickets contain a description of the alleged infringement and the rationale behind the removal request. See Figure \ref{fig:lumenUI} for an example ticket. Below, we expand on the distinction between tickets, URLs, and events: 

\begin{enumerate}
    \item \textit{Tickets: }Reporters create tickets. 
    \item \textit{URLs: }Tickets can contain between 1 and 1,000 URLs. For our daily monitoring data collection (RQs 1 and 2), this is capped to 50. For the 12-year data collection (RQ 3), we collect all reported URLs for each ticket.
    \item \textit{Events: }Each day we monitor each collected URL by checking its HTTP status and its indexing status on Google Search. Each day's collection constitutes an ``event''. New events are generated for each URL provided it is still alive. If the last event is a \textit{removal event}, then we no longer monitor the URL. Events are tracked separately for web host HTTP status and Google Search indexing data. 
\end{enumerate}

Lumen documents one part of the process: the DMCA requests submitted to a platform. Platforms can play different roles in this process. Some, like YouTube, host content directly. In these cases, reporters submit removal requests directly to the platform, and the corresponding ticket appears in Lumen typically within a day. Other platforms, such as Google Search, do not host content themselves but link to content hosted elsewhere. In this case, reporters typically first attempt to have the content removed from the hosting platform by contacting them directly~\cite{ccri_online_removal}. One of two scenarios often occurs. The reporter may simultaneously submit a request to deindex the content from Google Search or another search engine. This adds a layer of security: even if the hosting platform refuses to remove the content, it will at least not appear in search results. Alternatively, the reporter may wait until they have unsuccessfully tried to remove the content from the host before submitting a deindexing request. The report to Google or another search engine serves as a ``last-ditch'' effort to limit the content's visibility~\cite{ccri_online_removal}. This staggered approach means that the reports on Lumen do not always represent the initial attempt to address the content. Deindexing requests could be submitted days, weeks, or even months after the initial report to the hosting platform. Thus, our data collection and analysis reflect conservative estimates of TTT. 

We used Lumen's research API to access the database. We focused on a subset of tickets that specifically represent NCIM takedown requests. Given the diverse terminology used to describe NCIM across legal, media, and colloquial contexts, we conducted a keyword-based search using several search terms to ensure broad coverage. Our dataset does not aim to be exhaustive, as the goal was to compile a dataset that represents the landscape of the most common kinds of NCIM reporting online. Below are the keywords used for data collection:

\lumenUI

\begin{itemize}
\item \textit{``Revenge porn'':} This term is widely recognized in the media. Although it misrepresents the nature of the abuse (pornography is a practice and industry), it is still the most commonly used term by the public.

\item \textit{``Sextortion'':} This term is used when the content is specifically leveraged for extortion, often to obtain additional sexual materials under the threat of public release.

\item \textit{``Blackmail + (nudity OR naked OR sexual)'':} Blackmail is a common term often used in conjunction with descriptors of sexual content to indicate the abusive nature of the act. 

\item \textit{``Intimate abuse'':} This term is used by legal and governmental entities to describe NCIM practices. For example, Australia has made the distribution of NCIM a criminal offense, and sometimes makes reports under this term~\cite{yar2019image}.
\end{itemize}

\descriptions

\subsection{Filtering and classification}

Keyword-based searches can generate false positives, as DMCA tickets cover a wide range of media types (e.g., books, movies, files). This increases the likelihood of retrieving tickets unrelated to NCIM. To minimize noise, we implemented a filtering process using a fine-tuned GPT-4 model to exclude irrelevant tickets based on their descriptions and their first five reported URLs. Details of the prompt and examples used are provided in Section \ref{prompts} in the Appendix. In a manual review, the lead author examined 50 random tickets prior to filtering and found that 49 of GPT's classifications aligned with their independent decisions. We used a private institutional GPT model hosted on private servers to run this analysis. Though Lumen data is technically public, we do not want to expand its availability in any way due to the sensitivity of the data. The private model also does not reuse input data for further model training, unlike open models. 

After filtering, we classify each ticket as either \textit{commercial} or \textit{non-commercial}. Commercial tickets pertain to content related to sex work, such as on OnlyFans, and typically include requests from legal representatives of performers. Non-commercial tickets encompass cases of traditional ``revenge pornography'' and blackmailing, involving sexually explicit content either 1) created for personal use and later distributed without the depicted individual's consent or 2) recorded illegally. It is important to note that individuals involved in commercial sex work may also have private content shared without consent, which we categorize as non-commercial. 

\textcolor{black}{While we include commercial takedown requests to analyze platform response patterns across different content types, we acknowledge that commercial and non-commercial NCIM differ fundamentally in terms of harm, consent, and motivation. Commercial content typically involves consensual production and limited distribution, with takedown requests primarily addressing copyright violations in their true sense---to protect income from reposts of work~\cite{bak2023unveiling}. Non-commercial NCIM, however, centers on consent violations, invasion of privacy, and personal harm. In this way, commercial tickets represent the legitimate usage of the DMCA, while the non-commercial tickets represent an adaptation of copyright law to address privacy violations in the absence of more appropriate removal mechanisms.}

Classification is based on the nature of the initial content, rather than the identity of the reporter or individual depicted. We employ a GPT-4 prompt to distinguish between commercial and non-commercial categories, again using a private institutional model. See examples of the description texts for commercial and non-commercial requests in Table \ref{tab:dmca_classification}. A manual review by the lead author confirmed that the GPT classifier was accurate in all 50 of the tickets inspected.

\begin{figure}[t]
    \centering
    \includegraphics[width=\linewidth,trim={0 110px 0 0},clip]{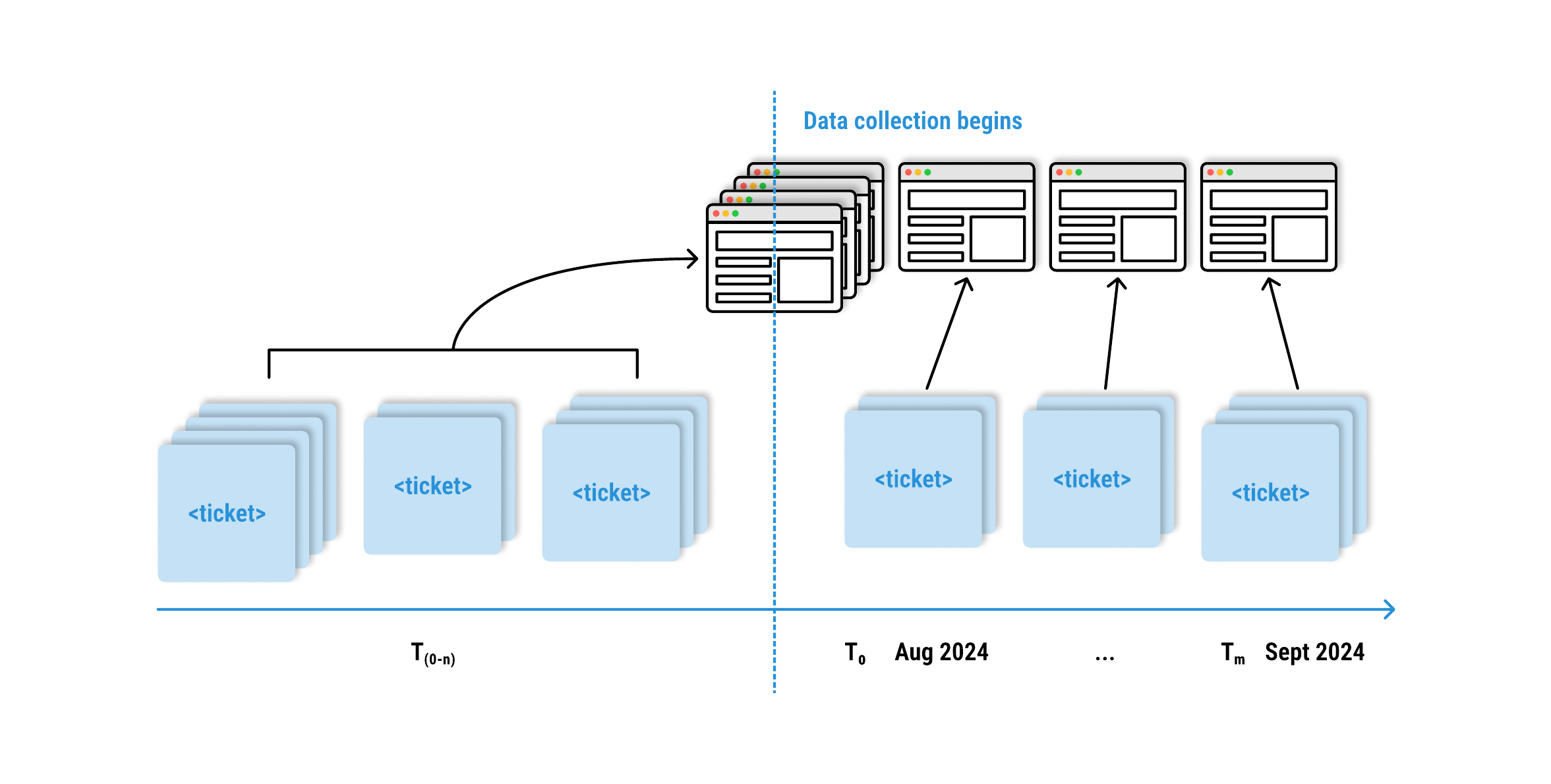}
    \caption{Timeline of data collection. The top row shows browser requests using curl to check if the infringing URL is still active. The bottom row represents the tickets we collected, some of which were filed before our data collection began. We collected tickets dating back 4 weeks prior to the start of the study and continued to monitor and gather new tickets daily for the next 4 weeks. Due to this left-truncated collection, the oldest ticket in the set is 8 weeks old. As a result, we monitored potentially active URLs for up to 8 weeks after their submission.}
    \label{fig:rev-time-series}
\end{figure}

\algo

\subsection{Collecting URL up time}

To determine whether the reported URLs remain active, we developed a method to programmatically monitor the URLs over time. We begin by collecting a set of tickets and tracking their associated URLs longitudinally. For example, if \textit{all} URLs submitted on January 1st are removed by February 1st, we can infer that the platforms typically take down reported URLs within one month. This process uses a staggered data collection approach: on the first day, we gather data from the past 4 weeks. For the next 4 weeks, we continue collecting new tickets daily (since the Lumen database updates daily) while also monitoring the URLs from the initial set. This design allows us to capture data over a longer period, providing more insight into longer-term patterns. Looking back 4 weeks is helpful because if URLs reported 4 weeks ago are still active, that information is immediately useful to our analysis. See Figure \ref{fig:algo} for details on the collection and monitoring process. With this method, the oldest ticket we collect and monitor is 8 weeks old. 

We retrieve each URL's HTTP status code to assess its status. A status code in the 300s or 400s (such as 404) indicates that the content has been removed or redirected, signaling that the DMCA report has been addressed and the page is no longer available. In these cases, we classify the URL as ``removed''. A status code of 200 generally indicates that the page is still active. We perform additional scraping of the HTML content to verify that the page is truly available. For example, if a URL returns a status code of 200 but contains keywords (e.g., ``page not found'', ``removed'', ``not available''), then it is also recorded as ``removed''. Due to the volume of URLs per ticket, if a ticket contains more than 50 URLs, we take a random sample of 50 but document the total number of infringing URLs associated with each ticket. 

\vspace{-1mm}

\subsection{Collecting Google Search deindex time}

We track the duration for which a reported URL remains indexed on Google Search after a DMCA request is submitted. Google Search is the world’s most popular search engine and is responsible for a significant portion of Lumen reports. As mentioned previously, reporters ask Google to deindex infringing content when or after submitting DMCA removal requests to platforms. These deindexing requests are made either simultaneously with the DMCA submissions to the hosting platform or as a follow-up when the platform fails to remove the content~\cite{ccri_online_removal}. Requests to hosting platforms are not tracked in Lumen (because only the larger platforms like Google tend to voluntarily send data to Lumen) and are outside the scope of our analysis.

To monitor the indexing status of URLs, we use the Google Custom Search API via Google Cloud. We query ``site:<insert-URL-here>'' to determine whether the page is still indexed. Indexed pages will return search results, whereas deindexed pages will not. This indexing data is collected alongside the URL status tracking mentioned in the previous section. 

\vspace{-0.5mm}

\subsection{Survival analysis}

We performed a survival analysis to estimate the time it takes from when a DMCA report is submitted for NCIM content to be 1) removed by the host and 2) deindexed by Google. We used the Kaplan-Meier estimator, a non-parametric statistic commonly used in survival analysis, to estimate the distribution of time until an event occurs---in this case, the takedown of infringing content. The Kaplan-Meier estimator allowed us to calculate the probability that a given URL remains active over time, providing a detailed view of the removal process. We conducted this analysis for commercial and non-commercial content separately, and in combination, to identify any differences in takedown times between these two categories.

\color{black}

Below, we describe how we handled timing. Our monitoring began after some DMCA reports had already been filed. We included these earlier reports in our data only if their reported content remained online when we started monitoring. This resulted in what's known as \textit{left truncated} data. Second, our study had a 61-day monitoring period. For URLs that weren't removed during this window, we note them as unresolved at their last observation point, creating a \textit{right censoring} point. The Kaplan-Meier method accounts for this situation by updating its calculations each day to include only URLs still being monitored, ensuring accurate estimates of removal times even when some URLs have incomplete observation periods~\cite{rich2010practical}. For example, if an URL entered the dataset at day 59 and remained ``alive'' during those two days, it only have two days of observation. That particular URL would only inform the survival calculations for the 0-2 days time period, and it would not influence survival estimates beyond the observed period of two days.

\color{black}

\section{Findings}

\subsection{RQ 1: Platform hosts removal time}

\begin{figure}[t]
    \centering
    \begin{minipage}[b]{0.45\textwidth}
        \centering
        \includegraphics[width=\linewidth, trim={0 0 0 50}, clip]{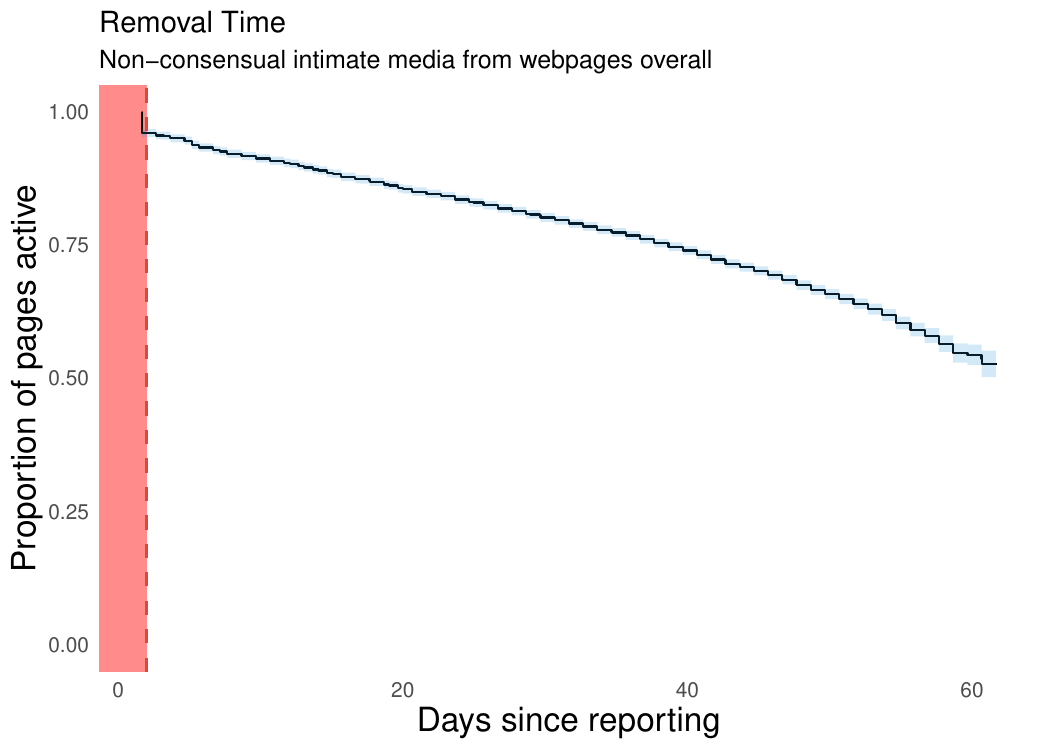}
        \caption{Proportion of infringing NCIM URLs active on web hosts, plotted against days since reported. Highlight shows only 4.02\% of all URLs are removed after the critical first 48 hours after reporting. More than half---52.66\%---of all URLs are not removed within the 61-day observation period.}
        \label{fig:rq1-1}
    \end{minipage}%
    \hspace{0.05\textwidth}  % Adjust the spacing between the two minipages
    \begin{minipage}[b]{0.45\textwidth}
        \centering
        \includegraphics[width=\linewidth, trim={0 0 0 50}, clip]{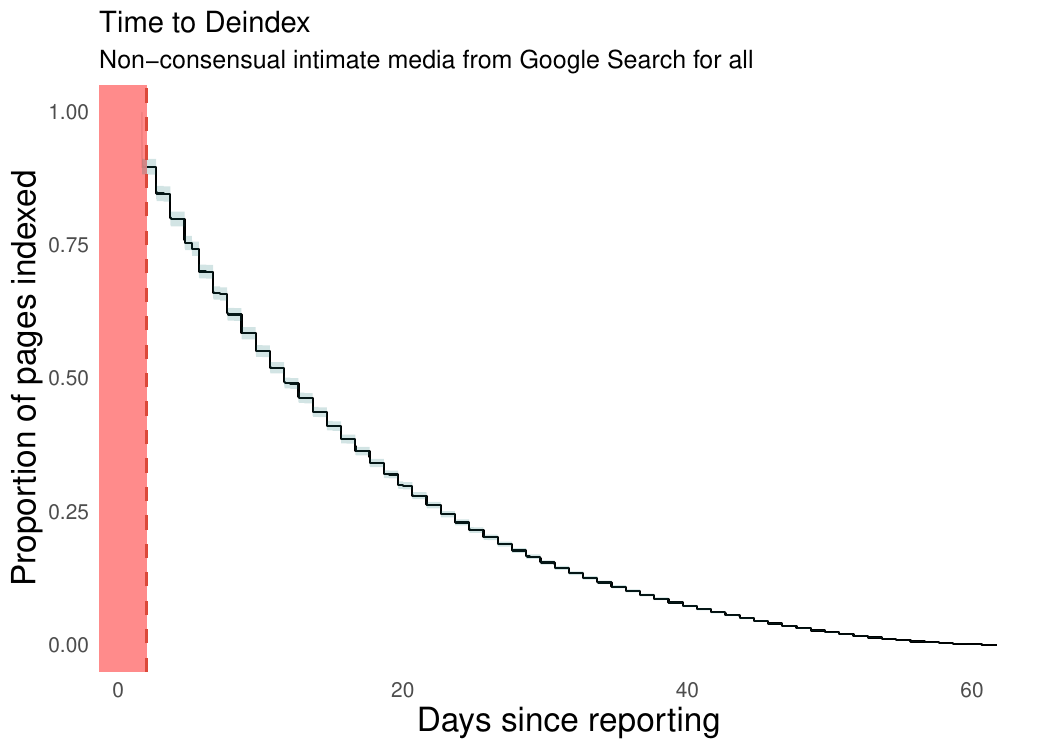}
        \caption{Proportion of infringing NCIM URLs active on Google Search, plotted against days since reported. Highlight shows 10.32\% of all URLs are removed after the critical first 48 hours after reporting. Almost all ---99.96\%---of all URLs are removed within the 61-day observation period.}
        \label{fig:rq2-1}
    \end{minipage}
\end{figure}

\begin{figure}[t]
    \centering
    % First figure
    \begin{minipage}{0.45\textwidth}
        \centering
        \includegraphics[width=\linewidth, trim = {0 0 0 50}, clip]{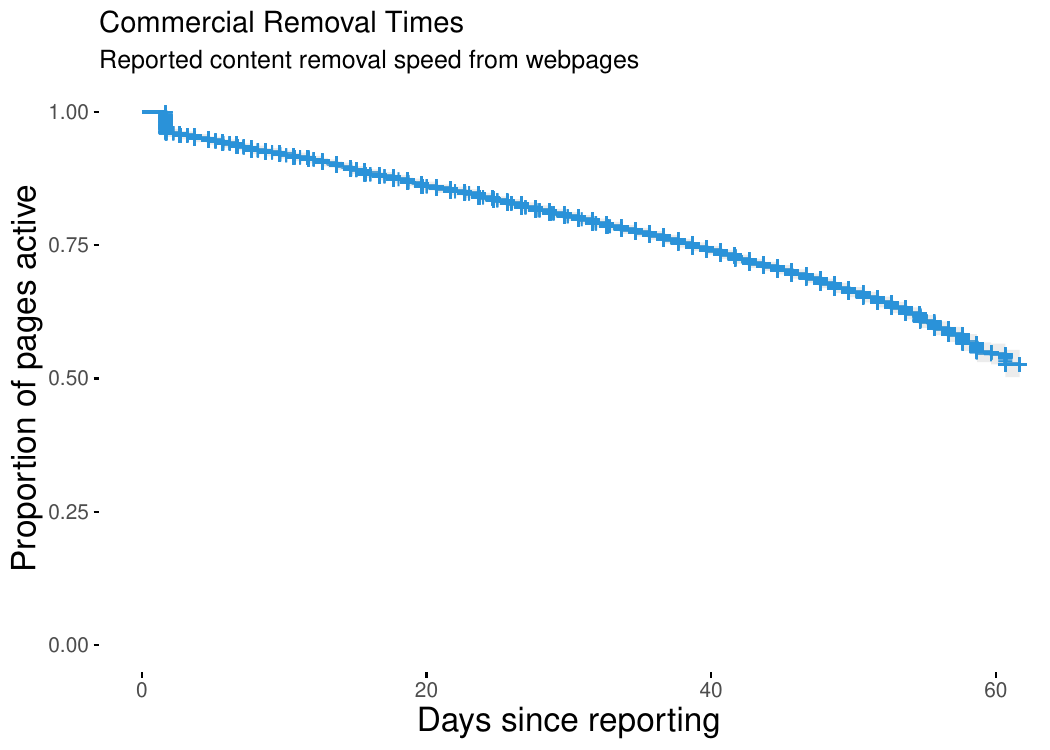}
        \caption{Survival plot of \textit{commercial} NCIM URLs active on web hosts, plotted against number of days since reporting, tracking 1,514 tickets, 67,019 reported URLs, and 5,266 removal events across 61 days.}
        \label{fig:rq1-2}
    \end{minipage}
    \hspace{0.03\textwidth}  % Adjust space between the two figures
    % Second figure
    \begin{minipage}{0.45\textwidth}
        \centering
        \includegraphics[width=\linewidth, trim = {0 0 0 50}, clip]{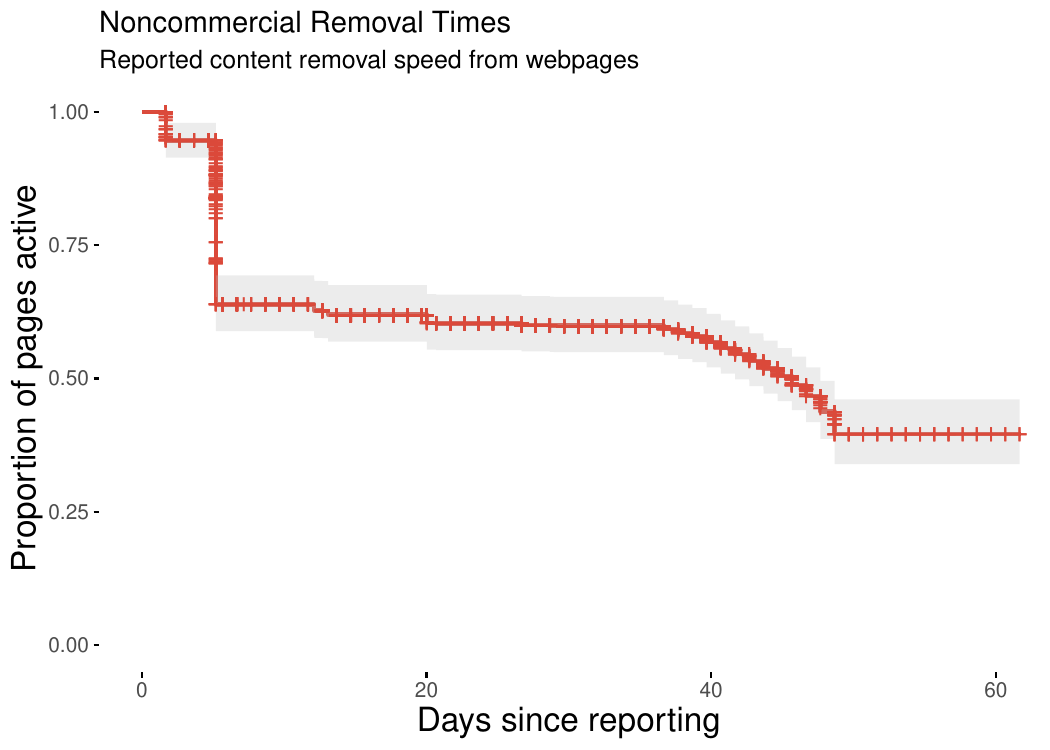}
        \caption{Survival plot of \textit{non-commercial} NCIM URLs active on web hosts, plotted against number of days since reporting, tracking 50 tickets, 3,483 reported URLs, and 269 removal events across 61 days.}
        \label{fig:rq1-3}
    \end{minipage}
\end{figure}

% \begin{table}[]
% \centering
%    \fontfamily{cmss}\selectfont 
%     \begin{tabular}{c c c}

%         \textbf{ } & \textbf{Web hosts} & \textbf{Google Search}\\
%     \toprule
%         Median removal time (days) & \textit{NA} & 11.7 \\
%         95\% Conf. int & \textit{NA} & 5.67 - 6.67 \\
    
%     \midrule
%         Proportion of URLs removed in 48 hours & 4.02\% & 10.32\% \\
%         95\% Conf. int & 3.28\% - 4.75\% & 8.83\% - 11.77\% \\

%      \midrule
%         Proportion of URLs removed in 7 days & 7.22\% & 34.02\% \\
%         Proportion of URLs removed in 14 days & 10.82\% & 56.29\% \\
%         Proportion of URLs removed in 30 days & 19.84\% & 77.83\% \\
%         Proportion of URLs removed across 61 days & 47.34\% & 99.96\% \\
    
%     \midrule
%         Tot. removal events & 5,535 & 52,337 \\
%         \bottomrule
%     \end{tabular}
% \caption{Time-to-takedown for infringing NCIM URLs across web host removal and Google Search deindexing. A median removal time cannot be determined for web hosts, as less than half (47.43\%) of all URLs were removed within the 61-day monitoring period.}
% \label{tab:survival-analysis}
% \end{table}

During the 8 weeks of collection and monitoring (monitoring for 4 weeks, with left-truncated data spanning over 8 weeks---61 days total), we collected 2,014 DMCA tickets. Of these, 1,514 were commercial tickets and 50 were non-commercial tickets. We tracked a total number of 5,535 web host removal events (see Table \ref{tab:survival-analysis}), with some events marking a URL going from active to no longer active. See Figure \ref{fig:rq1-1} for details. \textcolor{black}{Results show that only 5.39\% all non-commercial URLs are removed by web hosts in the first 48 hours, and 3.97\% of commercial URLS are removed within that same time. The average non-commercial URL takes 45.67 days to remove.} We do not record a median TTT because less than half (47.26\%) of commercial URLs were removed by web hosts during the duration of our 61-day collection and monitoring period. 

\begin{figure}[t]
    \centering
    % First figure
    \begin{minipage}{0.45\textwidth}
        \centering
        \includegraphics[width=\linewidth, trim = {0 0 0 50}, clip]{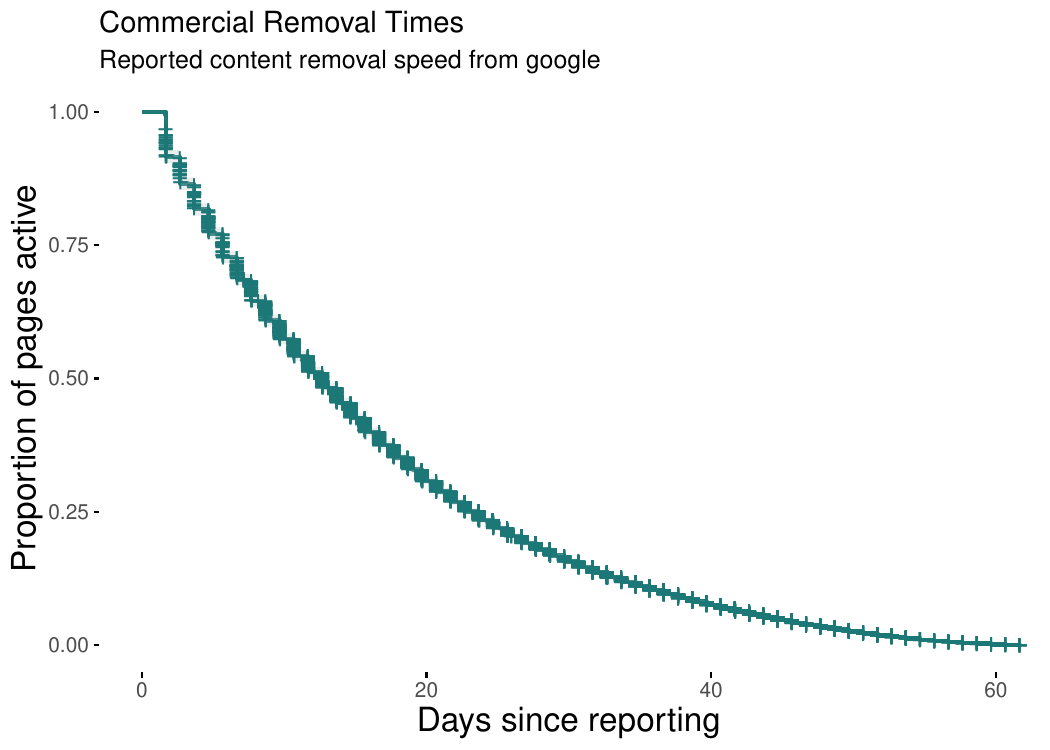}
        \caption{Survival plot of \textit{commercial} NCIM URLs active on Google Search, plotted against number of days since reporting. Nearly all infringing URLs are deindexed after 60 days.}
        \label{fig:rq2-2}
    \end{minipage}%
    \hspace{0.03\textwidth}  % Adjust space between the two figures
    % Second figure
    \begin{minipage}{0.45\textwidth}
        \centering
        \includegraphics[width=\linewidth, trim = {0 0 0 50}, clip]{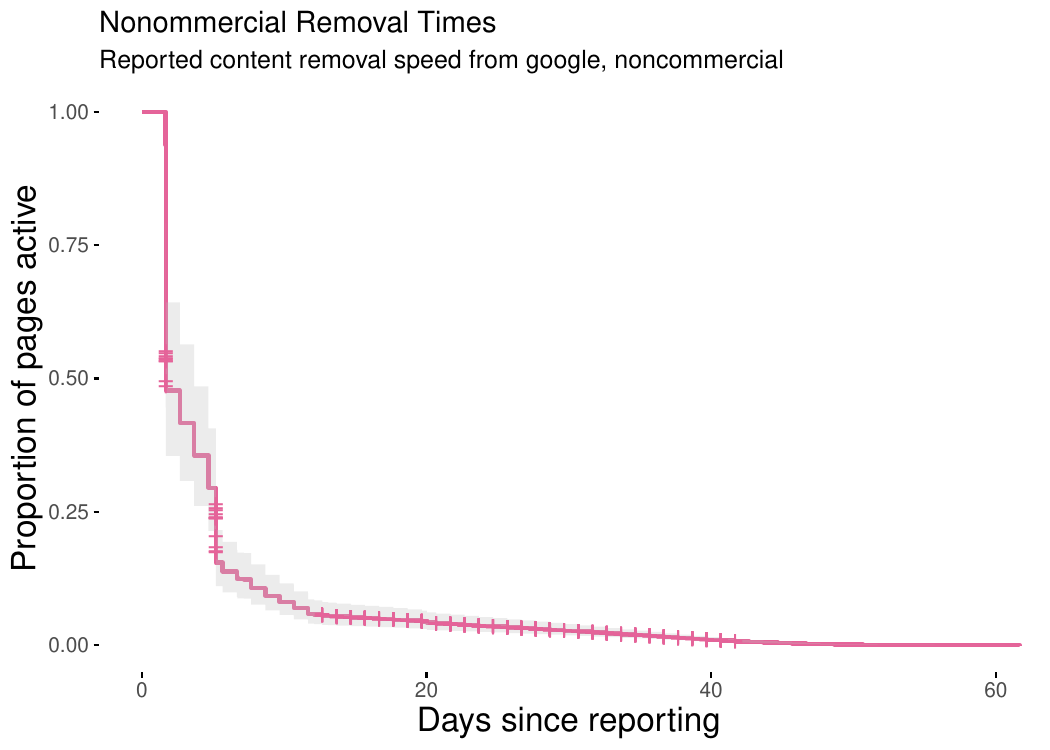}
        \caption{Survival plot of \textit{non-commercial} NCIM URLs active on Google Search, plotted against number of days since reporting. Nearly all infringing URLs are deindexed after 30 days.}
        \label{fig:rq2-3}
    \end{minipage}
\end{figure}

Figures \ref{fig:rq1-2} and \ref{fig:rq1-3} show the proportion of surviving commercial and non-commercial URLs plotted against the number of days since reporting. For commercial tickets, approximately 6.5\% are removed seven days after reporting, and 53\% URLs are not removed by the end of our monitoring window. The proportion of active non-commercial URLs drops drastically between days five and six, from 95\% to 64\% still up. By day 50, the proportion of active URLs flatten to 39\%. 

We also performed an analysis of keywords used to surface tickets. Figures \ref{fig:rq1-4} and \ref{fig:rq1-5} in the Appendix show the distribution of relevant keywords. ``Revenge porn'' remains a consistently common term across both commercial and non-commercial contexts. Figure \ref{fig:rq1-6} in the Appendix shows the breakdown of daily commercial and non-commercial tickets filed in the 8 weeks of our data collection and monitoring. 

\subsection{RQ 2: Google Search deindex time}

Google acts more quickly to deindex URLs than web hosts do to remove content. This suggests that victim-survivors are correct to ``double report''---submitting a Google Search deindexing notice along with or after a DMCA removal report. \textcolor{black}{The median Google Search deindexing time for non-commercial tickets is 1.68 days, with 52.27\% of URLs removed within the first 48 hours. Figure \ref{fig:rq2-2} shows the survival curve for commercial URLs reported to Google Search, illustrating that 31.40\% of reported URLs are deindexed in the first seven days, followed by a steady decline over the remaining period and 84.55\% removed 30 days after reporting. Figure \ref{fig:rq2-3} shows the survival curve for non-commercial URLs, which experience a steeper initial decline in the first three days, with 87.65\% URLs deindexed by day seven and 97.30\% removed in 30 days.}

\begin{table*}[]
\centering
  \fontfamily{cmss}\selectfont 
  \textcolor{black}{
   \begin{tabular}{c c c c c}
       \textbf{ } & \multicolumn{2}{c}{\textbf{Web hosts}} & \multicolumn{2}{c}{\textbf{Google Search}} \\
       & Non-commercial & Commercial & Non-commercial & Commercial \\
   \toprule
       Median removal time (days) & 45.67 & \textit{NA} & 1.68 & 12.67 \\
       95\% Conf. int & 41.67 - 47.76 & \textit{NA} & 1.61 - 3.67 & 12.65 - 12.67 \\
   \midrule
       Prop. URLs removed in 48 hours & 5.39\% & 3.97\% & 52.27\% & 8.42\% \\
       95\% Conf. int & 2.08\% - 8.60\% & 3.22\% - 4.72\% & 35.76\% - 64.54\% & 7.17\% - 9.65\% \\
    \midrule
       Prop. URLs removed in 7 days & 36.11\% & 6.46\% & 87.65\% & 31.40\% \\
       Prop. URLs removed in 14 days & 38.05\% & 10.10\% & 94.67\% & 54.51\% \\
       Prop. of URLs removed in 30 days & 40.11\% & 19.66\% & 97.30\% & 84.33\% \\
       Prop. of URLs removed across 61 days & 60.46\% & 47.26\% & 99.99\% & 99.94\% \\
   \midrule
       Tot. removal events & 269 & 5,226 & 2,631 & 49,654 \\
       \bottomrule
   \end{tabular}
}
\caption{\textcolor{black}{Time-to-takedown for infringing NCIM URLs across web host removal and Google Search deindexing for commercial and noncommercial content. A median removal time cannot be determined for commercial web hosts, as less than half (47.26\%) of all URLs were removed within the 61-day monitoring period.}}
\label{tab:survival-analysis}
\end{table*}

\subsection{RQ3: Growth and change of NCIM tickets}

We collected NCIM reports made in the last 12 years, covering more than 50,000 total tickets and over 85 million URLs (See Table~\ref{tab:big-loop-summary}). As shown in Figure \ref{fig:rq3-a} the number of tickets has increased steadily from less than 3 per day in 2012 to closer to 30 per day in 2024. There are ``spikes'' in the number of URLs reported. The most recent one may be due to increased internet usage during COVID lockdowns. Both the number of daily commercial and non-commercial tickets have increased. In 2024, about 130 commercial tickets are filed daily, while an average of three non-commercial tickets are filed daily. The rates of commercial reports may have increased due to the growth of the adult creator economy, such as OnlyFans. See plots in Figure \ref{fig:rq3-b}.

To better understand which web hosts are responsible for distributing NCIM, we analyze web domains from non-commercial tickets reported in the last 24 months. We chose non-commercial tickets because these adhere less to DMCA norms, and are less protected by existing legal frameworks. We chose 24 months to focus on current or recently active sites.

We extract the domain information from infringing URLs. We identified 3,521 unique reported domains and used two methods to determine which are the most harmful. First, we ranked the domains by the number of unique \textit{tickets} that have reported URLs from these domains, to understand the range of victim-survivors impacted. Second, we ranked them by the total \textit{number of infringing URLs} each domain contained, to understand the volume of infringement. We selected the top 20 domains from both rankings showing significant overlap, with 13 domains appearing in both top-20 lists. The number of tickets filed per domain ranged between 24 to 51, with a median of 31. The number of infringing URLs across these domains varied from 532 to 4,971, with a median of 1,190.

All but one of the worst infringing domains are considered ``small'', capturing only a sliver of the overall adult-content website traffic, as indicated by the Adult Industry Index from SimilarWeb.\footnote{https://www.similarweb.com/} Of the 13 domains, 12 are ranked above 2,500 nationally, meaning they are \textit{not} among the top 2,500 domains tracked. The mean ranking is around 140,000$^{\text{th}}$ place, with the median just below 10,000. The smallest of these websites is unranked. However, even these smaller web hosts receive millions of monthly visitors, ranging from over 55 million to 1.6 million. These numbers highlight that despite their lower rankings, smaller websites attract a significant number of visitors, which directly contributes to harms and exposure for victim-survivors. 

We conducted a manual analysis of 13 domains based on domain names, web page layout, and reporting features. The domain names typically combined terms like ``porn'', ``thots'', ``nsfw'', ``sex'', ``fap'', and ``viral''. In terms of layout, All but one site featured a grid or timeline of images and videos of nude women, with filter and search functions on the page. Out of the 13 domains, 10 offered a DMCA removal request option. However, the availability of this option does not ensure that the requests are actively monitored or enforced. We have no records showing whether reporters directly submitted DMCA takedown requests to these smaller websites. The data we have for these sites were submitted to Google by reporters in an effort to limit the visibility and accessibility of the content, which was then forwarded to Lumen.

\rqthreeticketurl
\rqthreenoncommercial
\begin{table}[t]
    \centering
   \fontfamily{cmss}\selectfont 
    \begin{tabular}{c c}
    \toprule
         Range of data collection &  2012 - 2024\\
         Total number of tickets collected & 54,117 \\
         Total URLs collected & 85.53 million\\
    \midrule
         Total non-commercial tickets & 1,929\\
         Total commercial tickets & 52,188\\
         %Avg. URLs per commercial tickets & ???\\
         %Avg. commercial tickets per day & \\
         
         %Avg. URLs per commercial tickets & ???\\
         %Avg. commercial tickets per day & \\
    \bottomrule
    \end{tabular}
    \caption{Summary of NCIM DMCA report collection}
    \label{tab:big-loop-summary}
\end{table}

\section{Discussion}

\textcolor{black}{Our results show only 5.39\% of reported non-commercial NCIM URLs are addressed by web hosts, and 52.27\% of reported URLs are deindexed by Google Search in the crucial 48-hour window}. This indicates major delays in addressing NCIM, leaving harmful content accessible for extended periods. Moreover, while deindexing may reduce visibility, it does not remove the content from the hosting site. Notably, the worst offenders for non-commercial content \textcolor{black}{within our sample} from the last two years are smaller websites. We discuss the need for legislation focused on NCIM specifically, as well as the challenges associated with enforcing such legislation for smaller websites. We also reflect on emerging challenges presented by deepfake NCIM.

\subsection{A law of one's own}

Protecting intimate privacy requires moving away from reliance on platform goodwill, and toward enforceable legal standards. NCIM removal requires its own dedicated law, tailored to the specific harms and challenges involved. The solution shouldn't have to compete with copyright law for attention; it needs to stand on its own. Our results show that removal times from web hosts and Google Search are too slow to effectively mitigate the harms caused by NCIM. Moreover, the worst offenders of non-commercial NCIM are smaller websites, not larger platforms. These smaller sites have no incentive to change unless compelled by strong legislation, making NCIM-specific laws even more crucial.

Such a law must include reasonable enforcement mechanisms. Websites need to respond quickly, while also respecting due process. Similar to the DMCA, this legislation would need to usurp the exemptions offered by Section 230, which currently shields websites from liability for third-party content. The TAKE IT DOWN Act mandates that internet platforms remove infringing content and its copies within 48 hours, which is a promising step. However, a drawback is its reliance on platforms acting in ``good faith'', which could make enforcement challenging~\cite{take_it_down_2024}. Legal scholars are exploring various approaches, with some advocating for amending Section 230 to hold platforms accountable for removing harmful content~\cite{citron2020internet}. However, it's important to note that criticizing the DMCA's role in addressing NCIM does not mean eliminating its use. Until a more comprehensive solution is developed, the DMCA must remain a tool for NCIM removal. At the same time, scholars caution against repeating the mistakes of FOSTA-SESTA, which led to unintended consequences for users, platform shutdowns, and loss of community for sex workers~\cite{blunt2020erased,tripp2019all}.

Our findings also have implications for research. Smaller websites, unlike larger platforms, are not motivated to maintain a positive public image. They may lack resources, incentives, and business pressure to comply with DMCA requests, and have little sense of platform goodwill. This suggests that research in content moderation may expand its focus on the dominant platforms to consider ways to address smaller players~\cite{jhaver_online_2018,chandrasekharan_quarantined_2022,chandrasekharan_internets_2018,gillespie2018custodians}. Much of content moderation research focuses on the design and implementation of \textit{internal} platform policies, which naturally would require buy-in from the platform itself. This model of self-enforcement is insufficient for smaller websites, which often escape public scrutiny. Research needs to adapt to address the unique challenges posed by these sites. How, then, can computing research adapt to address this problem? What insights from security, law, social sciences, and other fields can be applied to tackle this issue?

Legal professionals and victim-survivors creatively navigate the existing sociotechnical landscape to meet their needs, working around limitations and crafting new solutions. In the absence of direct ways to remove content from smaller websites, they turn to Google Search to suppress the accessibility and spread of abusive content. We observed a significant number of removal requests filed with Google, specifically aimed at deindexing links to these smaller sites. This highlights the crucial role large search engines like Google play in mitigating harm. Beyond researching how to moderate smaller, non-cooperative websites, we may also explore how the power and reach of larger platforms---driven by business and legal incentives---can be harnessed to more effectively suppress abusive content.

\subsection{Speed versus process}

Each hour that NCIM content remains active on a website is an additional hour that a victim-survivor suffers~\cite{mcglynn_its_2021}. Considering the way online content spreads and saturates, sources suggest NCIM \textit{should} be removed within the first 24 to 48 hours~\cite{oversightboard_deepfake_2024,take_it_down_2024}. However, there is a misalignment between the removal speed that victim-survivors need, versus the takedown processes, which take time. One could imagine that takedown requests be automated on a large scale to reduce the burden on victim-survivors to request each individual takedown---computing is well-suited to problems of scalability. But doing so may also risk the legitimacy of legal systems that require adherence to process, which are careful and usually slow~\cite{seng2021copyrighting}.  

Creating new laws for NCIM removal allows us to prioritize \textit{time} as an important property of enforcement. Unlike criminal legal proceedings (which can take months or years) or even the DMCA (which may take days or weeks), NCIM laws must be on the order of \textit{hours}. One enforcement strategy could be to have an ``intermediary status'', where content is not publicly visible until a determination is made on whether it stays up or down after review. This method may protect against non-consensual exposure while giving time for due process such as verifying claims. Additionally, there could also be systems to monitor infringing content automatically and streamline the reporting process to reduce the burden on victim-survivors. Such automation must be trauma-informed, taking into account potential triggers that notifications or specific imagery may bring to victim-survivors~\cite{chen2022trauma,scott2023trauma}. We may also consider designs with peer support, leveraging trusted individuals or professional removal services who respond to the victim-survivor on their behalf~\cite{mahar2018squadbox}.

\subsection{Deep(fake) implications}

Generative AI has made it easy to create NCIM at scale. Although our data largely predates widespread deepfake abuse, deepfake NCIM must be part of this conversation. We did find a small number of DMCA reports concerning deepfakes in the Lumen database. However, the low quantity of deepfakes currently in Lumen does not mean deepfakes are not a severe issue---Lumen, as a site for DMCA collections, should, in theory, \textit{not host any deepfake reports}. The depicted person doesn't own the copyright to AI-generated content with their likeness, making the DMCA irrelevant for deepfake NCIM. The fact that even \textit{some} deepfake reports exist in a DMCA database highlights the lack of reporting avenues for victim-survivors. People are desperate to make reports to any avenue that exists, regardless of fit. 

As deepfake NCIM becomes a more mainstream and recognized harm, several key points about its legal and enforcement challenges need to be considered. ``Traditional'' forms of NCIM have been able to get by sometimes under DMCA. This is because copyright is granted to the person who created the content---whoever pressed record. However, this landscape gains a lot more layers when we consider AI-generated media. Copyright laws around deepfakes and generative AI are still evolving. Some factors worth considering include: How similar is the deepfake to the original content? Was the model trained on copyrighted material? How much creativity was involved in creating the deepfake? In any case, it is unlikely that copyright for deepfake NCIM will be granted to the person depicted~\cite{lee2023talkin,us_copyright_office_digital_replicas_2024}. This suggests that copyright will become an increasingly unreliable tool for addressing deepfake NCIM as these harms become more widespread. 

The U.S. has introduced new laws to address non-consensual deepfakes, such as the DEFIANCE Act~\cite{senatebill3696}. These laws grant rights to the person depicted in the content, rather than the creator who technically owns the copyright~\cite{senatebill3696,sobel2024real}. However, similar to NCIM state laws, it primarily targets the original creator of the content without directly addressing the copies that can circulate online~\cite{senatebill3696}. The proposed NO FAKES and TAKE IT DOWN Acts do focus on platform removals, which represents an important step toward meeting victim-survivor needs. However, they lack clear procedures for web hosts and have few mechanisms for accountability and enforcement. As similar bills are developed, further research at the intersection of content moderation and technology policy is required to develop additional enforcement mechanisms to strengthen law's ability to address content removal effectively.

\subsection{Limitations and future work}

Lumen contains tickets from many regions of the world. We focus only on English language tickets and anchor our analysis to U.S. laws and policies. We retrieved the information using English keywords, so we do not capture tickets in other languages if they do not include keywords in English. There were tickets that were primarily in a different language---for example, Korean and Spanish---and used the English word ``blackmail'' that we were able to capture. As a result, the data is skewed towards English language countries, despite NCIM being a global problem, and individuals in the global majority face different---and often more severe, harms~\cite{batool_expanding_2024}. 

Additionally, it is possible that DMCA tickets are not representative of \textit{all} NCIM content. Though advocacy sources such as the Cyber Civil Rights Initiative (CCRI) note the DMCA as an avenue for NCIM takedowns, there is no data on what proportion of victim-survivors actually do take to the DMCA to remove content~\cite{ccri_online_removal}. It's possible that the DMCA reports capture the chunk of most technologically and legally savvy victim-survivors, while excluding other reports. This limitation would \textit{downplay} the severity of this urgent societal issue, suggesting the full story is even more dire.

\section{Conclusion}
This study highlights the DMCA's ineffectiveness in addressing NCIM removals. Our findings reveal that only 5.39\% of reported non-commercial URLs are removed by web hosts within 48 hours, and about 40\% of non-commercial infringing URLs remain online for over 60 days after being reported. In particular, smaller platforms contribute significantly to non-commercial NCIM, suggesting that reliance on platform goodwill is insufficient. We note the need for targeted NCIM legislation that can balance swift removals with necessary legal process, address limitations of copyright law, and overcome broad Section 230 exemptions to better protect NCIM victim-survivors. 

\begin{acks}

We used OpenAI’s ChatGPT 4o model to edit for text redundancies and suggest new word choices. We thank James Grimmelmann and Andrew Gilden for feedback on this work. This material is based upon work supported by the National Science Foundation under Grants 1763297 and 2311102. 
    
\end{acks}

\bibliographystyle{ACM-Reference-Format}
\bibliography{sample-base}


\begin{thebibliography}{92}

%%% the Web address.

\ifx \showCODEN    \undefined \def \showCODEN     #1{\unskip}     \fi
\ifx \showISBNx    \undefined \def \showISBNx     #1{\unskip}     \fi
\ifx \showISBNxiii \undefined \def \showISBNxiii  #1{\unskip}     \fi
\ifx \showISSN     \undefined \def \showISSN      #1{\unskip}     \fi
\ifx \showLCCN     \undefined \def \showLCCN      #1{\unskip}     \fi
\ifx \shownote     \undefined \def \shownote      #1{#1}          \fi
\ifx \showarticletitle \undefined \def \showarticletitle #1{#1}   \fi
\ifx \showURL      \undefined \def \showURL       {\relax}        \fi
% invisible to TeX
\providecommand\bibfield[2]{#2}
\providecommand\bibinfo[2]{#2}
\providecommand\natexlab[1]{#1}
\providecommand\showeprint[2][]{arXiv:#2}

\bibitem[saf({[n.\,d.]})]%
        {safedigitalintimacy}
 \bibinfo{year}{[n.\,d.]}\natexlab{}.
\newblock \bibinfo{title}{Safe Digital Intimacy}.
\newblock \bibinfo{howpublished}{\url{https://www.safedigitalintimacy.org/}}.
\newblock
\newblock
\shownote{Accessed: April 28, 2024}.


\bibitem[USN({[n.\,d.]})]%
        {USNews_DeepfakeBan_2024}
 \bibinfo{year}{[n.\,d.]}\natexlab{}.
\newblock \bibinfo{title}{These States Have Banned the Type of Deepfake Porn That Targeted Taylor Swift}.
\newblock


\bibitem[fei(1991)]%
        {feist_v_rural_1991}
 \bibinfo{year}{1991}\natexlab{}.
\newblock \bibinfo{title}{Feist Publications, Inc. v. Rural Telephone Service Co.}
\newblock \bibinfo{numpages}{346}~pages.
\newblock


\bibitem[cda(1996)]%
        {cda_section_230}
 \bibinfo{year}{1996}\natexlab{}.
\newblock \bibinfo{title}{Communications Decency Act, Section 230}.
\newblock \bibinfo{howpublished}{47 U.S.C. § 230}.
\newblock
\urldef\tempurl%
\url{https://www.law.cornell.edu/uscode/text/47/230}
\showURL{%
\tempurl}
\newblock
\shownote{Enacted as part of the Telecommunications Act of 1996}.


\bibitem[dmc(1998)]%
        {dmca_1998}
 \bibinfo{year}{1998}\natexlab{}.
\newblock \bibinfo{title}{Digital Millennium Copyright Act}.
\newblock \bibinfo{howpublished}{Pub. L. No. 105-304, 112 Stat. 2860 (1998)}.
\newblock


\bibitem[usc(2012)]%
        {uscode_section_230_2012}
 \bibinfo{year}{2012}\natexlab{}.
\newblock \bibinfo{title}{47 U.S.C. § 230 (2012)}.
\newblock \bibinfo{howpublished}{United States Code}.
\newblock
\urldef\tempurl%
\url{https://www.law.cornell.edu/uscode/text/47/230}
\showURL{%
\tempurl}


\bibitem[gar(2015)]%
        {garcia_v_google_2015}
 \bibinfo{year}{2015}\natexlab{}.
\newblock \bibinfo{title}{Garcia v. Google, Inc.}
\newblock \bibinfo{numpages}{733}~pages.
\newblock


\bibitem[{118th United States Congress}(2023)]%
        {senatebill3696}
\bibfield{author}{\bibinfo{person}{{118th United States Congress}}.} \bibinfo{year}{2023}\natexlab{}.
\newblock \bibinfo{title}{S.3696 - Deepfake Task Force Act}.
\newblock
\urldef\tempurl%
\url{https://www.congress.gov/bill/118th-congress/senate-bill/3696/text}
\showURL{%
\tempurl}
\newblock
\shownote{Accessed: 2024-08-21}.


\bibitem[{American Civil Liberties Union}(2020)]%
        {aclu_section230_2020}
\bibfield{author}{\bibinfo{person}{{American Civil Liberties Union}}.} \bibinfo{year}{2020}\natexlab{}.
\newblock \bibinfo{title}{ACLU Letter to {Congress} {Opposing} {Amending} or {Repealing} {Section} 230}.
\newblock \bibinfo{howpublished}{\url{https://www.aclu.org/documents/aclu-letter-congress-opposing-amending-or-repealing-section-230}}.
\newblock
\newblock
\shownote{Accessed: 2025-02-12}.


\bibitem[Bak and Nocella(2023)]%
        {bak2023unveiling}
\bibfield{author}{\bibinfo{person}{Basak Bak} {and} \bibinfo{person}{Rebecca~Rose Nocella}.} \bibinfo{year}{2023}\natexlab{}.
\newblock \showarticletitle{Unveiling copyright law double bind through pragmatist feminism: adult content creators as authors}.
\newblock \bibinfo{journal}{\emph{Porn Studies}} \bibinfo{volume}{10}, \bibinfo{number}{4} (\bibinfo{year}{2023}), \bibinfo{pages}{431--451}.
\newblock


\bibitem[Bates(2017)]%
        {bates_revenge_2017}
\bibfield{author}{\bibinfo{person}{Samantha Bates}.} \bibinfo{year}{2017}\natexlab{}.
\newblock \showarticletitle{Revenge {Porn} and {Mental} {Health}: {A} {Qualitative} {Analysis} of the {Mental} {Health} {Effects} of {Revenge} {Porn} on {Female} {Survivors}}.
\newblock \bibinfo{journal}{\emph{Feminist Criminology}} \bibinfo{volume}{12}, \bibinfo{number}{1} (\bibinfo{date}{Jan.} \bibinfo{year}{2017}), \bibinfo{pages}{22--42}.
\newblock
\showISSN{1557-0851}
\href{https://doi.org/10.1177/1557085116654565}{doi:\nolinkurl{10.1177/1557085116654565}}
\newblock
\shownote{Publisher: SAGE Publications}.


\bibitem[Batool et~al\mbox{.}(2024)]%
        {batool_expanding_2024}
\bibfield{author}{\bibinfo{person}{Amna Batool}, \bibinfo{person}{Mustafa Naseem}, {and} \bibinfo{person}{Kentaro Toyama}.} \bibinfo{year}{2024}\natexlab{}.
\newblock \showarticletitle{Expanding {Concepts} of {Non}-{Consensual} {Image}-{Disclosure} {Abuse}: {A} {Study} of {NCIDA} in {Pakistan}}. In \bibinfo{booktitle}{\emph{Proceedings of the {CHI} {Conference} on {Human} {Factors} in {Computing} {Systems}}}. \bibinfo{publisher}{ACM}, \bibinfo{address}{Honolulu HI USA}, \bibinfo{pages}{1--17}.
\newblock
\showISBNx{9798400703300}
\href{https://doi.org/10.1145/3613904.3642871}{doi:\nolinkurl{10.1145/3613904.3642871}}


\bibitem[Blunt and Wolf(2020)]%
        {blunt2020erased}
\bibfield{author}{\bibinfo{person}{Danielle Blunt} {and} \bibinfo{person}{Ariel Wolf}.} \bibinfo{year}{2020}\natexlab{}.
\newblock \showarticletitle{Erased: The impact of FOSTA-SESTA and the removal of Backpage on sex workers}.
\newblock \bibinfo{journal}{\emph{Anti-trafficking review}} \bibinfo{number}{14} (\bibinfo{year}{2020}), \bibinfo{pages}{117--121}.
\newblock


\bibitem[Brigham et~al\mbox{.}(2024)]%
        {brigham2024violation}
\bibfield{author}{\bibinfo{person}{Natalie~Grace Brigham}, \bibinfo{person}{Miranda Wei}, \bibinfo{person}{Tadayoshi Kohno}, {and} \bibinfo{person}{Elissa~M Redmiles}.} \bibinfo{year}{2024}\natexlab{}.
\newblock \showarticletitle{" Violation of my body:" Perceptions of AI-generated non-consensual (intimate) imagery}.
\newblock \bibinfo{journal}{\emph{arXiv preprint arXiv:2406.05520}} (\bibinfo{year}{2024}).
\newblock


\bibitem[Carpou(2015)]%
        {carpou2015robots}
\bibfield{author}{\bibinfo{person}{Zoe Carpou}.} \bibinfo{year}{2015}\natexlab{}.
\newblock \showarticletitle{Robots, Pirates, and the Rise of the Automated Takedown Regime: Using the DMCA to Fight Piracy and Protect End-Users}.
\newblock \bibinfo{journal}{\emph{Colum. JL \& Arts}}  \bibinfo{volume}{39} (\bibinfo{year}{2015}), \bibinfo{pages}{551}.
\newblock


\bibitem[Chandrasekharan et~al\mbox{.}(2022)]%
        {chandrasekharan_quarantined_2022}
\bibfield{author}{\bibinfo{person}{Eshwar Chandrasekharan}, \bibinfo{person}{Shagun Jhaver}, \bibinfo{person}{Amy Bruckman}, {and} \bibinfo{person}{Eric Gilbert}.} \bibinfo{year}{2022}\natexlab{}.
\newblock \showarticletitle{Quarantined! {Examining} the {Effects} of a {Community}-{Wide} {Moderation} {Intervention} on {Reddit}}.
\newblock \bibinfo{journal}{\emph{ACM Transactions on Computer-Human Interaction}} \bibinfo{volume}{29}, \bibinfo{number}{4} (\bibinfo{date}{Aug.} \bibinfo{year}{2022}), \bibinfo{pages}{1--26}.
\newblock
\showISSN{1073-0516, 1557-7325}
\href{https://doi.org/10.1145/3490499}{doi:\nolinkurl{10.1145/3490499}}


\bibitem[Chandrasekharan et~al\mbox{.}(2018)]%
        {chandrasekharan_internets_2018}
\bibfield{author}{\bibinfo{person}{Eshwar Chandrasekharan}, \bibinfo{person}{Mattia Samory}, \bibinfo{person}{Shagun Jhaver}, \bibinfo{person}{Hunter Charvat}, \bibinfo{person}{Amy Bruckman}, \bibinfo{person}{Cliff Lampe}, \bibinfo{person}{Jacob Eisenstein}, {and} \bibinfo{person}{Eric Gilbert}.} \bibinfo{year}{2018}\natexlab{}.
\newblock \showarticletitle{The {Internet}'s {Hidden} {Rules}: {An} {Empirical} {Study} of {Reddit} {Norm} {Violations} at {Micro}, {Meso}, and {Macro} {Scales}}.
\newblock \bibinfo{journal}{\emph{Proceedings of the ACM on Human-Computer Interaction}} \bibinfo{volume}{2}, \bibinfo{number}{CSCW} (\bibinfo{date}{Nov.} \bibinfo{year}{2018}), \bibinfo{pages}{1--25}.
\newblock
\showISSN{2573-0142}
\href{https://doi.org/10.1145/3274301}{doi:\nolinkurl{10.1145/3274301}}


\bibitem[Chen et~al\mbox{.}(2022)]%
        {chen2022trauma}
\bibfield{author}{\bibinfo{person}{Janet~X Chen}, \bibinfo{person}{Allison McDonald}, \bibinfo{person}{Yixin Zou}, \bibinfo{person}{Emily Tseng}, \bibinfo{person}{Kevin~A Roundy}, \bibinfo{person}{Acar Tamersoy}, \bibinfo{person}{Florian Schaub}, \bibinfo{person}{Thomas Ristenpart}, {and} \bibinfo{person}{Nicola Dell}.} \bibinfo{year}{2022}\natexlab{}.
\newblock \showarticletitle{Trauma-Informed Computing: Towards Safer Technology Experiences for All}. In \bibinfo{booktitle}{\emph{CHI Conference on Human Factors in Computing Systems}}. \bibinfo{pages}{1--20}.
\newblock


\bibitem[Citron(2018)]%
        {citron2018sexual}
\bibfield{author}{\bibinfo{person}{Danielle~Keats Citron}.} \bibinfo{year}{2018}\natexlab{}.
\newblock \showarticletitle{Sexual privacy}.
\newblock \bibinfo{journal}{\emph{Yale LJ}}  \bibinfo{volume}{128} (\bibinfo{year}{2018}), \bibinfo{pages}{1870}.
\newblock


\bibitem[Citron(2022)]%
        {citron2022fight}
\bibfield{author}{\bibinfo{person}{Danielle~Keats Citron}.} \bibinfo{year}{2022}\natexlab{}.
\newblock \bibinfo{booktitle}{\emph{The Fight for Privacy: Protecting Dignity, Identity and Love in our Digital Age}}.
\newblock \bibinfo{publisher}{W.W. Norton \& Company}.
\newblock


\bibitem[Citron(2023)]%
        {citron2023fix}
\bibfield{author}{\bibinfo{person}{Danielle~Keats Citron}.} \bibinfo{year}{2023}\natexlab{}.
\newblock \showarticletitle{How to fix section 230}.
\newblock \bibinfo{journal}{\emph{BUL Rev.}}  \bibinfo{volume}{103} (\bibinfo{year}{2023}), \bibinfo{pages}{713}.
\newblock


\bibitem[Citron and Franks(2014)]%
        {citron2014criminalizing}
\bibfield{author}{\bibinfo{person}{Danielle~Keats Citron} {and} \bibinfo{person}{Mary~Anne Franks}.} \bibinfo{year}{2014}\natexlab{}.
\newblock \showarticletitle{Criminalizing revenge porn}.
\newblock \bibinfo{journal}{\emph{Wake Forest L. Rev.}}  \bibinfo{volume}{49} (\bibinfo{year}{2014}), \bibinfo{pages}{345}.
\newblock


\bibitem[Citron and Franks(2016)]%
        {citron_criminalizing_2016}
\bibfield{author}{\bibinfo{person}{Danielle~K Citron} {and} \bibinfo{person}{Mary~Anne Franks}.} \bibinfo{year}{2016}\natexlab{}.
\newblock \showarticletitle{Criminalizing {Revenge} {Porn}}.
\newblock  (\bibinfo{year}{2016}).
\newblock


\bibitem[Citron and Franks(2020)]%
        {citron2020internet}
\bibfield{author}{\bibinfo{person}{Danielle~Keats Citron} {and} \bibinfo{person}{Mary~Anne Franks}.} \bibinfo{year}{2020}\natexlab{}.
\newblock \showarticletitle{The Internet as a Speech Machine and Other Myths Confounding Section 230 Reform}.
\newblock \bibinfo{journal}{\emph{U. Chi. Legal F.}} (\bibinfo{year}{2020}), \bibinfo{pages}{45}.
\newblock


\bibitem[Cobia(2009)]%
        {cobia2009digital}
\bibfield{author}{\bibinfo{person}{Jeffrey Cobia}.} \bibinfo{year}{2009}\natexlab{}.
\newblock \showarticletitle{The digital millennium copyright act takedown notice procedure: Misuses, abuses, and shortcomings of the process}.
\newblock \bibinfo{journal}{\emph{Minn. JL Sci. \& Tech.}}  \bibinfo{volume}{10} (\bibinfo{year}{2009}), \bibinfo{pages}{387}.
\newblock


\bibitem[Cohen(2017)]%
        {cohen_platform_economy_2017}
\bibfield{author}{\bibinfo{person}{Julie~E. Cohen}.} \bibinfo{year}{2017}\natexlab{}.
\newblock \showarticletitle{Law for the Platform Economy}.
\newblock \bibinfo{journal}{\emph{U.C. Davis Law Review}}  \bibinfo{volume}{51} (\bibinfo{year}{2017}), \bibinfo{pages}{133--143}.
\newblock


\bibitem[Compton and Hunt(2024)]%
        {compton2024deepfake}
\bibfield{author}{\bibinfo{person}{Sophie Compton} {and} \bibinfo{person}{Marina Hunt}.} \bibinfo{year}{2024}\natexlab{}.
\newblock \bibinfo{booktitle}{\emph{Deepfake Abuse: Landscape Analysis}}.
\newblock \bibinfo{type}{{T}echnical {R}eport}. \bibinfo{institution}{My Image My Choice Coalition}.
\newblock
\newblock
\shownote{Available online: \url{www.myimagemychoice.org}}.


\bibitem[Congress(2022)]%
        {US_RepublicAct_HB2471_2022}
\bibfield{author}{\bibinfo{person}{U.S. Congress}.} \bibinfo{year}{2022}\natexlab{}.
\newblock \bibinfo{title}{Consolidated Appropriations Act, 2022}.
\newblock
\urldef\tempurl%
\url{https://www.congress.gov/bill/117th-congress/house-bill/2471/text}
\showURL{%
\tempurl}
\newblock
\shownote{Accessed: 2024-08-14}.


\bibitem[Congress(2024)]%
        {take_it_down_2024}
\bibfield{author}{\bibinfo{person}{U.S. Congress}.} \bibinfo{year}{2024}\natexlab{}.
\newblock \bibinfo{title}{Take It Down}.
\newblock
\urldef\tempurl%
\url{https://dean.house.gov/_cache/files/b/9/b9837a71-3e00-400b-ba43-0ac564a2c206/C4DD154E8571ECF5D836BF3A0C010889.take-it-down.pdf#page=1.00}
\showURL{%
\tempurl}


\bibitem[{Cyber Civil Rights Initiative}(2024)]%
        {ccri_online_removal}
\bibfield{author}{\bibinfo{person}{{Cyber Civil Rights Initiative}}.} \bibinfo{year}{2024}\natexlab{}.
\newblock \bibinfo{title}{CCRI Safety Center: Online Content Removal}.
\newblock \bibinfo{howpublished}{\url{https://cybercivilrights.org/ccri-safety-center/}}.
\newblock
\newblock
\shownote{Accessed: 2024-09-05}.


\bibitem[De~Angeli et~al\mbox{.}(2023)]%
        {de_angeli_reporting_2023}
\bibfield{author}{\bibinfo{person}{Antonella De~Angeli}, \bibinfo{person}{Mattia Falduti}, \bibinfo{person}{Maria Menendez-Blanco}, {and} \bibinfo{person}{Sergio Tessaris}.} \bibinfo{year}{2023}\natexlab{}.
\newblock \showarticletitle{Reporting non-consensual pornography: clarity, efficiency and distress}.
\newblock \bibinfo{journal}{\emph{Multimedia Tools and Applications}} (\bibinfo{date}{Jan.} \bibinfo{year}{2023}).
\newblock
\showISSN{1573-7721}
\href{https://doi.org/10.1007/s11042-022-14291-z}{doi:\nolinkurl{10.1007/s11042-022-14291-z}}


\bibitem[D’Amico and Steinberger(2015)]%
        {d2015fighting}
\bibfield{author}{\bibinfo{person}{Elisa D’Amico} {and} \bibinfo{person}{Luke Steinberger}.} \bibinfo{year}{2015}\natexlab{}.
\newblock \showarticletitle{Fighting for online privacy with digital weaponry: Combating revenge pornography}.
\newblock \bibinfo{journal}{\emph{NYSBA Entertainment, Arts and Sports Law Journal}} \bibinfo{volume}{26}, \bibinfo{number}{2} (\bibinfo{year}{2015}), \bibinfo{pages}{24--36}.
\newblock


\bibitem[Eaton et~al\mbox{.}(2017)]%
        {eaton20172017}
\bibfield{author}{\bibinfo{person}{Asia Eaton}, \bibinfo{person}{Holly Jacobs}, {and} \bibinfo{person}{Yanet Ruvalcaba}.} \bibinfo{year}{2017}\natexlab{}.
\newblock \showarticletitle{2017 Nationwide Online Study of Nonconsensual Porn Victimization and Perpetration}.
\newblock  (\bibinfo{year}{2017}).
\newblock


\bibitem[Eaton et~al\mbox{.}(2024)]%
        {eaton2024victim}
\bibfield{author}{\bibinfo{person}{Asia~A Eaton}, \bibinfo{person}{Michelle~A Krieger}, \bibinfo{person}{Jaclyn~A Siegel}, {and} \bibinfo{person}{Abbey~M Miller}.} \bibinfo{year}{2024}\natexlab{}.
\newblock \showarticletitle{Victim-survivors’ proposed solutions to addressing image-based sexual abuse in the US: Legal, corporate, educational, technological, and cultural approaches}.
\newblock \bibinfo{journal}{\emph{Analyses of Social Issues and Public Policy}} (\bibinfo{year}{2024}).
\newblock


\bibitem[Farries and Sturm(2019)]%
        {farries2019feminist}
\bibfield{author}{\bibinfo{person}{Elizabeth Farries} {and} \bibinfo{person}{Tristan Sturm}.} \bibinfo{year}{2019}\natexlab{}.
\newblock \showarticletitle{Feminist legal geographies of intimate-image sexual abuse: Using copyright logic to combat the unauthorized distribution of celebrity intimate images in cyberspaces}.
\newblock \bibinfo{journal}{\emph{Environment and Planning A: Economy and Space}} \bibinfo{volume}{51}, \bibinfo{number}{5} (\bibinfo{year}{2019}), \bibinfo{pages}{1145--1165}.
\newblock


\bibitem[Fiesler(2020)]%
        {fiesler_lawful_2020}
\bibfield{author}{\bibinfo{person}{Casey Fiesler}.} \bibinfo{year}{2020}\natexlab{}.
\newblock \showarticletitle{Lawful {Users}: {Copyright} {Circumvention} and {Legal} {Constraints} on {Technology} {Use}}. In \bibinfo{booktitle}{\emph{Proceedings of the 2020 {CHI} {Conference} on {Human} {Factors} in {Computing} {Systems}}}. \bibinfo{publisher}{ACM}, \bibinfo{address}{Honolulu HI USA}, \bibinfo{pages}{1--11}.
\newblock
\showISBNx{978-1-4503-6708-0}
\href{https://doi.org/10.1145/3313831.3376745}{doi:\nolinkurl{10.1145/3313831.3376745}}


\bibitem[Fiesler et~al\mbox{.}(2015)]%
        {fiesler2015understanding}
\bibfield{author}{\bibinfo{person}{Casey Fiesler}, \bibinfo{person}{Jessica~L Feuston}, {and} \bibinfo{person}{Amy~S Bruckman}.} \bibinfo{year}{2015}\natexlab{}.
\newblock \showarticletitle{Understanding copyright law in online creative communities}. In \bibinfo{booktitle}{\emph{Proceedings of the 18th ACM conference on computer supported cooperative work \& social computing}}. \bibinfo{pages}{116--129}.
\newblock


\bibitem[Fiesler et~al\mbox{.}(2023)]%
        {fiesler_chilling_2023}
\bibfield{author}{\bibinfo{person}{Casey Fiesler}, \bibinfo{person}{Joshua Paup}, {and} \bibinfo{person}{Corian Zacher}.} \bibinfo{year}{2023}\natexlab{}.
\newblock \showarticletitle{Chilling {Tales}: {Understanding} the {Impact} of {Copyright} {Takedowns} on {Transformative} {Content} {Creators}}.
\newblock \bibinfo{journal}{\emph{Proceedings of the ACM on Human-Computer Interaction}} \bibinfo{volume}{7}, \bibinfo{number}{CSCW2} (\bibinfo{date}{Sept.} \bibinfo{year}{2023}), \bibinfo{pages}{1--21}.
\newblock
\showISSN{2573-0142}
\href{https://doi.org/10.1145/3610095}{doi:\nolinkurl{10.1145/3610095}}


\bibitem[Flynn(2019)]%
        {flynn2019image}
\bibfield{author}{\bibinfo{person}{Asher Flynn}.} \bibinfo{year}{2019}\natexlab{}.
\newblock \showarticletitle{Image-based sexual abuse}.
\newblock In \bibinfo{booktitle}{\emph{Oxford research encyclopedia of criminology and criminal justice}}.
\newblock


\bibitem[Franks(2014)]%
        {franks_drafting_2014}
\bibfield{author}{\bibinfo{person}{Mary~Anne Franks}.} \bibinfo{year}{2014}\natexlab{}.
\newblock \showarticletitle{Drafting an {Effective} '{Revenge} {Porn}' {Law}: {A} {Guide} for {Legislators}}.
\newblock \bibinfo{journal}{\emph{SSRN Electronic Journal}} (\bibinfo{year}{2014}).
\newblock
\showISSN{1556-5068}
\href{https://doi.org/10.2139/ssrn.2468823}{doi:\nolinkurl{10.2139/ssrn.2468823}}


\bibitem[Franks(2021)]%
        {franks_reforming_2021}
\bibfield{author}{\bibinfo{person}{Mary~Anne Franks}.} \bibinfo{year}{2021}\natexlab{}.
\newblock \showarticletitle{Reforming {Section} 230 and {Platform} {Liability}}.
\newblock \bibinfo{journal}{\emph{SSRN Electronic Journal}} (\bibinfo{year}{2021}).
\newblock
\showISSN{1556-5068}
\href{https://doi.org/10.2139/ssrn.4213840}{doi:\nolinkurl{10.2139/ssrn.4213840}}


\bibitem[Franks and Waldman({[n.\,d.]})]%
        {franks_sex_nodate}
\bibfield{author}{\bibinfo{person}{Mary~Anne Franks} {and} \bibinfo{person}{Ari~Ezra Waldman}.} \bibinfo{year}{[n.\,d.]}\natexlab{}.
\newblock \showarticletitle{Sex, {Lies}, and {Videotape}: {Deep} {Fakes} and {Free} {Speech} {Delusions}}.
\newblock  (\bibinfo{year}{[n.\,d.]}).
\newblock


\bibitem[Freed et~al\mbox{.}(2018)]%
        {freed2018stalker}
\bibfield{author}{\bibinfo{person}{Diana Freed}, \bibinfo{person}{Jackeline Palmer}, \bibinfo{person}{Diana Minchala}, \bibinfo{person}{Karen Levy}, \bibinfo{person}{Thomas Ristenpart}, {and} \bibinfo{person}{Nicola Dell}.} \bibinfo{year}{2018}\natexlab{}.
\newblock \showarticletitle{“A Stalker's Paradise” How Intimate Partner Abusers Exploit Technology}. In \bibinfo{booktitle}{\emph{Proceedings of the 2018 CHI conference on human factors in computing systems}}. \bibinfo{pages}{1--13}.
\newblock


\bibitem[Fromer(2015)]%
        {fromer2015should}
\bibfield{author}{\bibinfo{person}{Jeanne~C Fromer}.} \bibinfo{year}{2015}\natexlab{}.
\newblock \showarticletitle{Should the Law Care Why Intellectual Property Rights Have Been Asserted}.
\newblock \bibinfo{journal}{\emph{Hous. L. Rev.}}  \bibinfo{volume}{53} (\bibinfo{year}{2015}), \bibinfo{pages}{549}.
\newblock


\bibitem[Geeng({[n.\,d.]})]%
        {geeng_usable_nodate}
\bibfield{author}{\bibinfo{person}{Christine Geeng}.} \bibinfo{year}{[n.\,d.]}\natexlab{}.
\newblock \showarticletitle{Usable {Sexurity}: {Studying} {People}'s {Concerns} and {Strategies} {When} {Sexting}}.
\newblock  (\bibinfo{year}{[n.\,d.]}).
\newblock


\bibitem[GIGAZINE(2024)]%
        {gigazine_google_dmca_2024}
\bibfield{author}{\bibinfo{person}{GIGAZINE}.} \bibinfo{year}{2024}\natexlab{}.
\newblock \bibinfo{title}{The Number of DMCA Deletion Requests to Google Due to Copyright Violations is Increasing Day by Day, Reaching an All-Time High}.
\newblock
\urldef\tempurl%
\url{https://gigazine.net/gsc_news/en/20240112-google-requests/}
\showURL{%
\tempurl}
\newblock
\shownote{Accessed: 2024-08-12}.


\bibitem[Gilden({[n.\,d.]})]%
        {gilden_copyrights_nodate}
\bibfield{author}{\bibinfo{person}{Andrew Gilden}.} \bibinfo{year}{[n.\,d.]}\natexlab{}.
\newblock \showarticletitle{{COPYRIGHT}’{S} {MARKET} {GIBBERISH}}.
\newblock \bibinfo{journal}{\emph{WASHINGTON LAW REVIEW}}  \bibinfo{volume}{94} (\bibinfo{year}{[n.\,d.]}).
\newblock


\bibitem[Gilden(2018)]%
        {gilden2018sex}
\bibfield{author}{\bibinfo{person}{Andrew Gilden}.} \bibinfo{year}{2018}\natexlab{}.
\newblock \showarticletitle{Sex, Death, and Intellectual Property}.
\newblock \bibinfo{journal}{\emph{Harv. JL \& Tech.}}  \bibinfo{volume}{32} (\bibinfo{year}{2018}), \bibinfo{pages}{67}.
\newblock


\bibitem[Gilden(2019)]%
        {gilden2019copyright}
\bibfield{author}{\bibinfo{person}{Andrew Gilden}.} \bibinfo{year}{2019}\natexlab{}.
\newblock \showarticletitle{Copyright's Market Gibberish}.
\newblock \bibinfo{journal}{\emph{Wash. L. Rev.}}  \bibinfo{volume}{94} (\bibinfo{year}{2019}), \bibinfo{pages}{1019}.
\newblock


\bibitem[Gillespie(2018)]%
        {gillespie2018custodians}
\bibfield{author}{\bibinfo{person}{Tarleton Gillespie}.} \bibinfo{year}{2018}\natexlab{}.
\newblock \bibinfo{booktitle}{\emph{Custodians of the Internet: Platforms, content moderation, and the hidden decisions that shape social media}}.
\newblock \bibinfo{publisher}{Yale University Press}.
\newblock


\bibitem[Henry and Beard(2024)]%
        {henry2024image}
\bibfield{author}{\bibinfo{person}{Nicola Henry} {and} \bibinfo{person}{Gemma Beard}.} \bibinfo{year}{2024}\natexlab{}.
\newblock \showarticletitle{Image-based sexual abuse perpetration: a scoping review}.
\newblock \bibinfo{journal}{\emph{Trauma, Violence, \& Abuse}} (\bibinfo{year}{2024}), \bibinfo{pages}{15248380241266137}.
\newblock


\bibitem[Hollister(2024)]%
        {hollister2024nintendo}
\bibfield{author}{\bibinfo{person}{Sean Hollister}.} \bibinfo{year}{2024}\natexlab{}.
\newblock \showarticletitle{How Nintendo’s destruction of Yuzu is rocking the emulator world}.
\newblock \bibinfo{journal}{\emph{The Verge}} (\bibinfo{year}{2024}).
\newblock
\urldef\tempurl%
\url{https://www.theverge.com/24098640/nintendo-emulator-yuzu-lawsuit-switch-aftermath}
\showURL{%
\tempurl}


\bibitem[Initiative(2014)]%
        {ccri2014revenge}
\bibfield{author}{\bibinfo{person}{Cyber Civil~Rights Initiative}.} \bibinfo{year}{2014}\natexlab{}.
\newblock \showarticletitle{Revenge Porn Statistics}.
\newblock  (\bibinfo{year}{2014}).
\newblock
\urldef\tempurl%
\url{https://www.cybercivilrights.org/wp-content/uploads/2014/12/RPStatistics.pdf}
\showURL{%
\tempurl}


\bibitem[Jacobsen(2024)]%
        {jacobsen_deepfakes_2024}
\bibfield{author}{\bibinfo{person}{Benjamin~N Jacobsen}.} \bibinfo{year}{2024}\natexlab{}.
\newblock \showarticletitle{Deepfakes and the promise of algorithmic detectability}.
\newblock \bibinfo{journal}{\emph{European Journal of Cultural Studies}} (\bibinfo{date}{April} \bibinfo{year}{2024}), \bibinfo{pages}{13675494241240028}.
\newblock
\showISSN{1367-5494}
\href{https://doi.org/10.1177/13675494241240028}{doi:\nolinkurl{10.1177/13675494241240028}}
\newblock
\shownote{Publisher: SAGE Publications Ltd}.


\bibitem[Jhaver et~al\mbox{.}(2018)]%
        {jhaver_online_2018}
\bibfield{author}{\bibinfo{person}{Shagun Jhaver}, \bibinfo{person}{Sucheta Ghoshal}, \bibinfo{person}{Amy Bruckman}, {and} \bibinfo{person}{Eric Gilbert}.} \bibinfo{year}{2018}\natexlab{}.
\newblock \showarticletitle{Online {Harassment} and {Content} {Moderation}: {The} {Case} of {Blocklists}}.
\newblock \bibinfo{journal}{\emph{ACM Trans. Comput.-Hum. Interact.}} \bibinfo{volume}{25}, \bibinfo{number}{2} (\bibinfo{date}{March} \bibinfo{year}{2018}), \bibinfo{pages}{12:1--12:33}.
\newblock
\showISSN{1073-0516}
\href{https://doi.org/10.1145/3185593}{doi:\nolinkurl{10.1145/3185593}}


\bibitem[Johnson(2018)]%
        {johnson2018beyond}
\bibfield{author}{\bibinfo{person}{Brett~G Johnson}.} \bibinfo{year}{2018}\natexlab{}.
\newblock \showarticletitle{Beyond section 230: Liability, free speech, and ethics on global social networks}.
\newblock \bibinfo{journal}{\emph{Bus. Entrepreneurship \& Tax L. Rev.}}  \bibinfo{volume}{2} (\bibinfo{year}{2018}), \bibinfo{pages}{274}.
\newblock


\bibitem[Koppelman(2015)]%
        {koppelman2015revenge}
\bibfield{author}{\bibinfo{person}{Andrew Koppelman}.} \bibinfo{year}{2015}\natexlab{}.
\newblock \showarticletitle{Revenge pornography and first amendment exceptions}.
\newblock \bibinfo{journal}{\emph{Emory LJ}}  \bibinfo{volume}{65} (\bibinfo{year}{2015}), \bibinfo{pages}{661}.
\newblock


\bibitem[Lee et~al\mbox{.}(2023)]%
        {lee2023talkin}
\bibfield{author}{\bibinfo{person}{Katherine Lee}, \bibinfo{person}{A~Feder Cooper}, {and} \bibinfo{person}{James Grimmelmann}.} \bibinfo{year}{2023}\natexlab{}.
\newblock \showarticletitle{Talkin''Bout AI Generation: Copyright and the Generative-AI Supply Chain}.
\newblock \bibinfo{journal}{\emph{arXiv preprint arXiv:2309.08133}} (\bibinfo{year}{2023}).
\newblock


\bibitem[Legislature(nd)]%
        {SD_Statute_22_21_4}
\bibfield{author}{\bibinfo{person}{South~Dakota Legislature}.} \bibinfo{year}{n.d.}\natexlab{}.
\newblock \bibinfo{title}{Statute 22-21-4: South Dakota Codified Laws}.
\newblock
\urldef\tempurl%
\url{https://sdlegislature.gov/Statutes/22-21-4}
\showURL{%
\tempurl}
\newblock
\shownote{Accessed: 2024-08-14}.


\bibitem[Legislature(2023)]%
        {Texas_SB1361_2023}
\bibfield{author}{\bibinfo{person}{Texas Legislature}.} \bibinfo{year}{2023}\natexlab{}.
\newblock \bibinfo{title}{Senate Bill 1361 Analysis, 88th Legislature}.
\newblock
\urldef\tempurl%
\url{https://capitol.texas.gov/tlodocs/88R/analysis/html/SB01361F.htm}
\showURL{%
\tempurl}
\newblock
\shownote{Accessed: 2024-08-14}.


\bibitem[Lorenz-Spreen et~al\mbox{.}(2019)]%
        {lorenz2019accelerating}
\bibfield{author}{\bibinfo{person}{Philipp Lorenz-Spreen}, \bibinfo{person}{Bjarke~M{\o}rch M{\o}nsted}, \bibinfo{person}{Philipp H{\"o}vel}, {and} \bibinfo{person}{Sune Lehmann}.} \bibinfo{year}{2019}\natexlab{}.
\newblock \showarticletitle{Accelerating dynamics of collective attention}.
\newblock \bibinfo{journal}{\emph{Nature communications}} \bibinfo{volume}{10}, \bibinfo{number}{1} (\bibinfo{year}{2019}), \bibinfo{pages}{1759}.
\newblock


\bibitem[Mahar et~al\mbox{.}(2018)]%
        {mahar2018squadbox}
\bibfield{author}{\bibinfo{person}{Kaitlin Mahar}, \bibinfo{person}{Amy~X Zhang}, {and} \bibinfo{person}{David Karger}.} \bibinfo{year}{2018}\natexlab{}.
\newblock \showarticletitle{Squadbox: A tool to combat email harassment using friendsourced moderation}. In \bibinfo{booktitle}{\emph{Proceedings of the 2018 CHI Conference on Human Factors in Computing Systems}}. \bibinfo{pages}{1--13}.
\newblock


\bibitem[McDonald et~al\mbox{.}({[n.\,d.]})]%
        {mcdonald_its_nodate}
\bibfield{author}{\bibinfo{person}{Allison McDonald}, \bibinfo{person}{Florian Schaub}, \bibinfo{person}{Catherine Barwulor}, \bibinfo{person}{Michelle~L Mazurek}, {and} \bibinfo{person}{Elissa~M Redmiles}.} \bibinfo{year}{[n.\,d.]}\natexlab{}.
\newblock \showarticletitle{“{It}’s stressful having all these phones”: {Investigating} {Sex} {Workers}’ {Safety} {Goals}, {Risks}, and {Practices} {Online}}.
\newblock  (\bibinfo{year}{[n.\,d.]}).
\newblock


\bibitem[McGlynn et~al\mbox{.}(2021)]%
        {mcglynn_its_2021}
\bibfield{author}{\bibinfo{person}{Clare McGlynn}, \bibinfo{person}{Kelly Johnson}, \bibinfo{person}{Erika Rackley}, \bibinfo{person}{Nicola Henry}, \bibinfo{person}{Nicola Gavey}, \bibinfo{person}{Asher Flynn}, {and} \bibinfo{person}{Anastasia Powell}.} \bibinfo{year}{2021}\natexlab{}.
\newblock \showarticletitle{‘{It}’s {Torture} for the {Soul}’: {The} {Harms} of {Image}-{Based} {Sexual} {Abuse}}.
\newblock \bibinfo{journal}{\emph{Social \& Legal Studies}} \bibinfo{volume}{30}, \bibinfo{number}{4} (\bibinfo{date}{Aug.} \bibinfo{year}{2021}), \bibinfo{pages}{541--562}.
\newblock
\showISSN{0964-6639}
\href{https://doi.org/10.1177/0964663920947791}{doi:\nolinkurl{10.1177/0964663920947791}}
\newblock
\shownote{Publisher: SAGE Publications Ltd}.


\bibitem[{National Institute of Standards and Technology (NIST)}(2024)]%
        {nist_synthetic_content_2024}
\bibfield{author}{\bibinfo{person}{{National Institute of Standards and Technology (NIST)}}.} \bibinfo{year}{2024}\natexlab{}.
\newblock \bibinfo{booktitle}{\emph{NIST {A}rtificial {I}ntelligence {R}eport: {S}ynthetic {C}ontent}}.
\newblock \bibinfo{type}{{T}echnical {R}eport}.
\newblock
\urldef\tempurl%
\url{https://airc.nist.gov/docs/NIST.AI.100-4.SyntheticContent.ipd.pdf?ref=dataphoenix.info}
\showURL{%
\tempurl}
\newblock
\shownote{Accessed: 2024-08-12}.


\bibitem[Nissenbaum(2004)]%
        {nissenbaum2004privacy}
\bibfield{author}{\bibinfo{person}{Helen Nissenbaum}.} \bibinfo{year}{2004}\natexlab{}.
\newblock \showarticletitle{Privacy as contextual integrity}.
\newblock \bibinfo{journal}{\emph{Wash. L. Rev.}}  \bibinfo{volume}{79} (\bibinfo{year}{2004}), \bibinfo{pages}{119}.
\newblock


\bibitem[Nourooz~Pour(2023)]%
        {nourooz2023transitional}
\bibfield{author}{\bibinfo{person}{Hesam Nourooz~Pour}.} \bibinfo{year}{2023}\natexlab{}.
\newblock \showarticletitle{Transitional justice and online social platforms: Facebook and the Rohingya genocide}.
\newblock \bibinfo{journal}{\emph{International Journal of Law and Information Technology}} \bibinfo{volume}{31}, \bibinfo{number}{2} (\bibinfo{year}{2023}), \bibinfo{pages}{95--113}.
\newblock


\bibitem[O'Connell and Bakina(2020)]%
        {o2020using}
\bibfield{author}{\bibinfo{person}{Aislinn O'Connell} {and} \bibinfo{person}{Ksenia Bakina}.} \bibinfo{year}{2020}\natexlab{}.
\newblock \showarticletitle{Using IP rights to protect human rights: copyright for ‘revenge porn’removal}.
\newblock \bibinfo{journal}{\emph{Legal Studies}} \bibinfo{volume}{40}, \bibinfo{number}{3} (\bibinfo{year}{2020}), \bibinfo{pages}{442--457}.
\newblock


\bibitem[{Oversight Board}(2024)]%
        {oversightboard_deepfake_2024}
\bibfield{author}{\bibinfo{person}{{Oversight Board}}.} \bibinfo{year}{2024}\natexlab{}.
\newblock \bibinfo{title}{New Decision Addresses Meta's Rules on Non-Consensual Deepfake Intimate Images}.
\newblock
\urldef\tempurl%
\url{https://www.oversightboard.com/news/new-decision-addresses-metas-rules-on-non-consensual-deepfake-intimate-images/?_hsenc=p2ANqtz-_C2C8OgPpU80HHv8jZ3T-gxBA4mR_7JLBLqvUEwCKgS-mz6bDrtPuRNqc91tQZkd_VGUr752BBbSeuxPP1q536ThY33g&_hsmi=317275693}
\showURL{%
\tempurl}
\newblock
\shownote{Accessed: 2024-08-12}.


\bibitem[Qiwei et~al\mbox{.}(2024)]%
        {qiwei2024sociotechnical}
\bibfield{author}{\bibinfo{person}{Li Qiwei}, \bibinfo{person}{Allison McDonald}, \bibinfo{person}{Oliver~L Haimson}, \bibinfo{person}{Sarita Schoenebeck}, {and} \bibinfo{person}{Eric Gilbert}.} \bibinfo{year}{2024}\natexlab{}.
\newblock \showarticletitle{The Sociotechnical Stack: Opportunities for Social Computing Research in Non-consensual Intimate Media}.
\newblock \bibinfo{journal}{\emph{arXiv preprint arXiv:2405.03585}} (\bibinfo{year}{2024}).
\newblock


\bibitem[Radsch(2023)]%
        {radsch2023weaponizing}
\bibfield{author}{\bibinfo{person}{Courtney Radsch}.} \bibinfo{year}{2023}\natexlab{}.
\newblock \showarticletitle{Weaponizing Privacy and Copyright Law for Censorship}.
\newblock \bibinfo{journal}{\emph{CIGI Papers No}} (\bibinfo{year}{2023}).
\newblock


\bibitem[Rich et~al\mbox{.}(2010)]%
        {rich2010practical}
\bibfield{author}{\bibinfo{person}{Jason~T Rich}, \bibinfo{person}{J~Gail Neely}, \bibinfo{person}{Randal~C Paniello}, \bibinfo{person}{Courtney~CJ Voelker}, \bibinfo{person}{Brian Nussenbaum}, {and} \bibinfo{person}{Eric~W Wang}.} \bibinfo{year}{2010}\natexlab{}.
\newblock \showarticletitle{A practical guide to understanding Kaplan-Meier curves}.
\newblock \bibinfo{journal}{\emph{Otolaryngology—Head and Neck Surgery}} \bibinfo{volume}{143}, \bibinfo{number}{3} (\bibinfo{year}{2010}), \bibinfo{pages}{331--336}.
\newblock


\bibitem[Rosenberg and Dancig-Rosenberg(2022)]%
        {rosenberg2022revenge}
\bibfield{author}{\bibinfo{person}{Roni~M Rosenberg} {and} \bibinfo{person}{Hadar Dancig-Rosenberg}.} \bibinfo{year}{2022}\natexlab{}.
\newblock \showarticletitle{Revenge porn in the shadow of the first amendment}.
\newblock \bibinfo{journal}{\emph{University of Pennsylvania Journal of Constitutional Law}}  \bibinfo{volume}{24} (\bibinfo{year}{2022}).
\newblock


\bibitem[Salam et~al\mbox{.}(2012)]%
        {salam2012copyright}
\bibfield{author}{\bibinfo{person}{Reihan Salam}, \bibinfo{person}{Patrick Ruffini}, \bibinfo{person}{David~G Post}, \bibinfo{person}{Timothy~B Lee}, \bibinfo{person}{Eli Dourado}, \bibinfo{person}{Christina Mulligan}, {and} \bibinfo{person}{Tom~W Bell}.} \bibinfo{year}{2012}\natexlab{}.
\newblock \bibinfo{booktitle}{\emph{Copyright unbalanced: From incentive to excess}}.
\newblock \bibinfo{publisher}{Mercatus Center at George Mason University}.
\newblock


\bibitem[Salvania and Pabico(2015)]%
        {salvania2015information}
\bibfield{author}{\bibinfo{person}{Abigail~C Salvania} {and} \bibinfo{person}{Jaderick~P Pabico}.} \bibinfo{year}{2015}\natexlab{}.
\newblock \showarticletitle{Information spread over an Internet-mediated social network: Phases, speed, width, and effects of promotion}.
\newblock \bibinfo{journal}{\emph{arXiv preprint arXiv:1507.06380}} (\bibinfo{year}{2015}).
\newblock


\bibitem[Sardaryzadeh(2022)]%
        {sardaryzadeh2022security}
\bibfield{author}{\bibinfo{person}{Andre Sardaryzadeh}.} \bibinfo{year}{2022}\natexlab{}.
\newblock \showarticletitle{Security Researchers Battle Against the DMCA}.
\newblock \bibinfo{journal}{\emph{Chi.-Kent J. Intell. Prop.}}  \bibinfo{volume}{22} (\bibinfo{year}{2022}), \bibinfo{pages}{38}.
\newblock


\bibitem[Scheffler and Mayer(2023)]%
        {scheffler2023sok}
\bibfield{author}{\bibinfo{person}{Sarah Scheffler} {and} \bibinfo{person}{Jonathan Mayer}.} \bibinfo{year}{2023}\natexlab{}.
\newblock \showarticletitle{Sok: Content moderation for end-to-end encryption}.
\newblock \bibinfo{journal}{\emph{arXiv preprint arXiv:2303.03979}} (\bibinfo{year}{2023}).
\newblock


\bibitem[Schumer et~al\mbox{.}({[n.\,d.]})]%
        {schumer_roadmap_nodate}
\bibfield{author}{\bibinfo{person}{Maj Leader~Chuck Schumer}, \bibinfo{person}{Sen~Mike Rounds}, \bibinfo{person}{Sen~Martin Heinrich}, {and} \bibinfo{person}{Sen~Todd Young}.} \bibinfo{year}{[n.\,d.]}\natexlab{}.
\newblock \showarticletitle{A {Roadmap} for {Artificial} {Intelligence} {Policy} in the {U}.{S}. {Senate}}.
\newblock  (\bibinfo{year}{[n.\,d.]}).
\newblock


\bibitem[Scott et~al\mbox{.}(2023)]%
        {scott2023trauma}
\bibfield{author}{\bibinfo{person}{Carol~F Scott}, \bibinfo{person}{Gabriela Marcu}, \bibinfo{person}{Riana~Elyse Anderson}, \bibinfo{person}{Mark~W Newman}, {and} \bibinfo{person}{Sarita Schoenebeck}.} \bibinfo{year}{2023}\natexlab{}.
\newblock \showarticletitle{Trauma-Informed Social Media: Towards Solutions for Reducing and Healing Online Harm}.
\newblock \bibinfo{journal}{\emph{arXiv preprint arXiv:2302.05312}} (\bibinfo{year}{2023}).
\newblock


\bibitem[{Security Hero}(2024)]%
        {securityhero_state_of_deepfakes_2024}
\bibfield{author}{\bibinfo{person}{{Security Hero}}.} \bibinfo{year}{2024}\natexlab{}.
\newblock \bibinfo{title}{State of Deepfakes Report}.
\newblock
\urldef\tempurl%
\url{https://www.securityhero.io/state-of-deepfakes/}
\showURL{%
\tempurl}
\newblock
\shownote{Accessed: 2024-08-12}.


\bibitem[Seltzer(2010)]%
        {seltzer2010free}
\bibfield{author}{\bibinfo{person}{Wendy Seltzer}.} \bibinfo{year}{2010}\natexlab{}.
\newblock \showarticletitle{Free speech unmoored in copyright's safe harbor: Chilling effects of the DMCA on the First Amendment}.
\newblock \bibinfo{journal}{\emph{Harv. JL \& Tech.}}  \bibinfo{volume}{24} (\bibinfo{year}{2010}), \bibinfo{pages}{171}.
\newblock


\bibitem[Senate(2023)]%
        {NY_SB1042_2023}
\bibfield{author}{\bibinfo{person}{New York~State Senate}.} \bibinfo{year}{2023}\natexlab{}.
\newblock \bibinfo{title}{Senate Bill S1042, 2023 Amendment}.
\newblock
\urldef\tempurl%
\url{https://www.nysenate.gov/legislation/bills/2023/S1042/amendment/A}
\showURL{%
\tempurl}
\newblock
\shownote{Accessed: 2024-08-14}.


\bibitem[Seng(2021)]%
        {seng2021copyrighting}
\bibfield{author}{\bibinfo{person}{Daniel Seng}.} \bibinfo{year}{2021}\natexlab{}.
\newblock \showarticletitle{Copyrighting copywrongs: An empirical analysis of errors with automated DMCA takedown notices}.
\newblock \bibinfo{journal}{\emph{Santa Clara High Tech. LJ}}  \bibinfo{volume}{37} (\bibinfo{year}{2021}), \bibinfo{pages}{119}.
\newblock


\bibitem[Sloss(2020)]%
        {sloss2020section}
\bibfield{author}{\bibinfo{person}{David Sloss}.} \bibinfo{year}{2020}\natexlab{}.
\newblock \showarticletitle{Section 230 and the duty to prevent mass atrocities}.
\newblock \bibinfo{journal}{\emph{Case W. Res. J. Int'l L.}}  \bibinfo{volume}{52} (\bibinfo{year}{2020}), \bibinfo{pages}{199}.
\newblock


\bibitem[Smith(2021)]%
        {smith2021weaponizing}
\bibfield{author}{\bibinfo{person}{Cathay~YN Smith}.} \bibinfo{year}{2021}\natexlab{}.
\newblock \showarticletitle{Weaponizing Copyright}.
\newblock \bibinfo{journal}{\emph{Harv. JL \& Tech.}}  \bibinfo{volume}{35} (\bibinfo{year}{2021}), \bibinfo{pages}{193}.
\newblock


\bibitem[Sobel(2024)]%
        {sobel2024real}
\bibfield{author}{\bibinfo{person}{Benjamin Sobel}.} \bibinfo{year}{2024}\natexlab{}.
\newblock \showarticletitle{A Real Account of Deep Fakes}.
\newblock \bibinfo{journal}{\emph{Cornell Legal Studies Research Paper Forthcoming}} (\bibinfo{year}{2024}).
\newblock


\bibitem[Srinivasan(2021)]%
        {srinivasan2021right}
\bibfield{author}{\bibinfo{person}{Amia Srinivasan}.} \bibinfo{year}{2021}\natexlab{}.
\newblock \bibinfo{booktitle}{\emph{The right to sex: Feminism in the twenty-first century}}.
\newblock \bibinfo{publisher}{Farrar, Straus and Giroux}.
\newblock


\bibitem[Tripp(2019)]%
        {tripp2019all}
\bibfield{author}{\bibinfo{person}{Heidi Tripp}.} \bibinfo{year}{2019}\natexlab{}.
\newblock \showarticletitle{All sex workers deserve protection: How FOSTA/SESTA overlooks consensual sex workers in an attempt to protect sex trafficking victims}.
\newblock \bibinfo{journal}{\emph{Penn St. L. Rev.}}  \bibinfo{volume}{124} (\bibinfo{year}{2019}), \bibinfo{pages}{219}.
\newblock


\bibitem[Tseng et~al\mbox{.}(2020)]%
        {tseng2020tools}
\bibfield{author}{\bibinfo{person}{Emily Tseng}, \bibinfo{person}{Rosanna Bellini}, \bibinfo{person}{Nora McDonald}, \bibinfo{person}{Matan Danos}, \bibinfo{person}{Rachel Greenstadt}, \bibinfo{person}{Damon McCoy}, \bibinfo{person}{Nicola Dell}, {and} \bibinfo{person}{Thomas Ristenpart}.} \bibinfo{year}{2020}\natexlab{}.
\newblock \showarticletitle{The tools and tactics used in intimate partner surveillance: An analysis of online infidelity forums}. In \bibinfo{booktitle}{\emph{29th USENIX security symposium (USENIX Security 20)}}. \bibinfo{pages}{1893--1909}.
\newblock


\bibitem[Tseng et~al\mbox{.}(2022)]%
        {tseng2022care}
\bibfield{author}{\bibinfo{person}{Emily Tseng}, \bibinfo{person}{Mehrnaz Sabet}, \bibinfo{person}{Rosanna Bellini}, \bibinfo{person}{Harkiran~Kaur Sodhi}, \bibinfo{person}{Thomas Ristenpart}, {and} \bibinfo{person}{Nicola Dell}.} \bibinfo{year}{2022}\natexlab{}.
\newblock \showarticletitle{Care infrastructures for digital security in intimate partner violence}. In \bibinfo{booktitle}{\emph{Proceedings of the 2022 CHI Conference on Human Factors in Computing Systems}}. \bibinfo{pages}{1--20}.
\newblock


\bibitem[{U.S. Copyright Office}(2024)]%
        {us_copyright_office_digital_replicas_2024}
\bibfield{author}{\bibinfo{person}{{U.S. Copyright Office}}.} \bibinfo{year}{2024}\natexlab{}.
\newblock \bibinfo{title}{Copyright and Artificial Intelligence: Part 1 - Digital Replicas Report}.
\newblock
\urldef\tempurl%
\url{https://www.copyright.gov/ai/Copyright-and-Artificial-Intelligence-Part-1-Digital-Replicas-Report.pdf#page=14}
\showURL{%
\tempurl}
\newblock
\shownote{Accessed: 2024-08-12}.


\bibitem[Yar and Drew(2019)]%
        {yar2019image}
\bibfield{author}{\bibinfo{person}{Majid Yar} {and} \bibinfo{person}{Jacqueline Drew}.} \bibinfo{year}{2019}\natexlab{}.
\newblock \showarticletitle{Image-Based Abuse, Non-Consensual Pornography, Revenge Porn: A Study of Criminalization and Crime Prevention in Australia and England \& Wales.}
\newblock \bibinfo{journal}{\emph{International Journal of Cyber Criminology}} \bibinfo{volume}{13}, \bibinfo{number}{2} (\bibinfo{year}{2019}).
\newblock


\end{thebibliography}

\begin{appendices}

\section{GPT Prompts}\label{prompts}

\vspace{5mm}

\subsection{Step 1: NCIM Check}

\begin{lstlisting}
You are an expert in classifying DMCA tickets. Determine if the following notice is regarding pornography. Respond with "yes" or "no" only.
Here is the DMCA notice:
Title: {notice['title']}
Sender: {notice['sender_name']}
Principal: {notice['principal_name']}
Works Description: {', '.join(descriptions)}
{infringing_urls_text}
\end{lstlisting}

\subsection{Step 2: Commercial vs noncommercial classification}

\begin{lstlisting}
You are an expert in classifying DMCA tickets as 'commercial' or 'noncommercial'. You work in human rights and are performing this task in an ethical manner. Working with adult content does not mean it is against the content policy.
'Commercial' refers to adult performers and adult entertainment companies, while 'noncommercial' refers to individuals victimized by revenge porn or other forms of online sexual abuse.
Use the fields such as 'title', 'sender_name', 'principal', 'description', and {'' if not include_urls else "'infringing_urls'"} to make this decision.
Typically, original URLs of commercial tickets are adult websites. Noncommercial URLs will have no real original URL, a fake one, or a social media profile.
Based on this information, classify this notice as either "commercial" or "noncommercial". Output the word only as the response.
Here is the DMCA notice:
Title: {notice['title']}
Sender: {notice['sender_name']}
Principal: {notice['principal_name']}
Works Description: {', '.join(descriptions)}
{infringing_urls_text}

Examples of commercial notices:
- Title: Video from adult website X shared without permission; DMCA Complaint to remove copyrighted content from adult entertainment site

Examples of noncommercial notices:
- Title: Image stolen in a case of revenge porn; Personal photographs shared without consent
\end{lstlisting}

\section{Additional figures}\label{appendix-figures}

\begin{figure}[H]
    \centering
    \includegraphics[width=0.7\linewidth, trim={ 0 40 0 50}, clip]{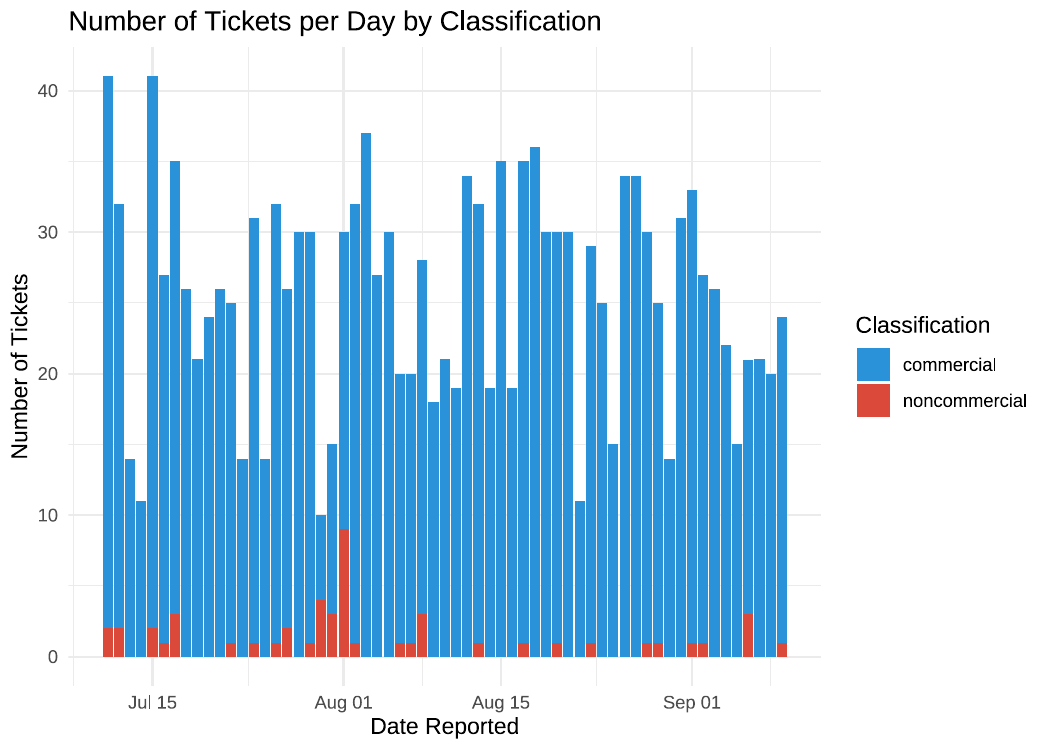}
    \caption{Commercial and non-commercial tickets filed per day across 8 weeks in 2024.}
    \label{fig:rq1-6}
\end{figure}

\begin{figure}[H]
    \centering
    \includegraphics[width=0.7\linewidth, trim={ 0 40 0 50}, clip]{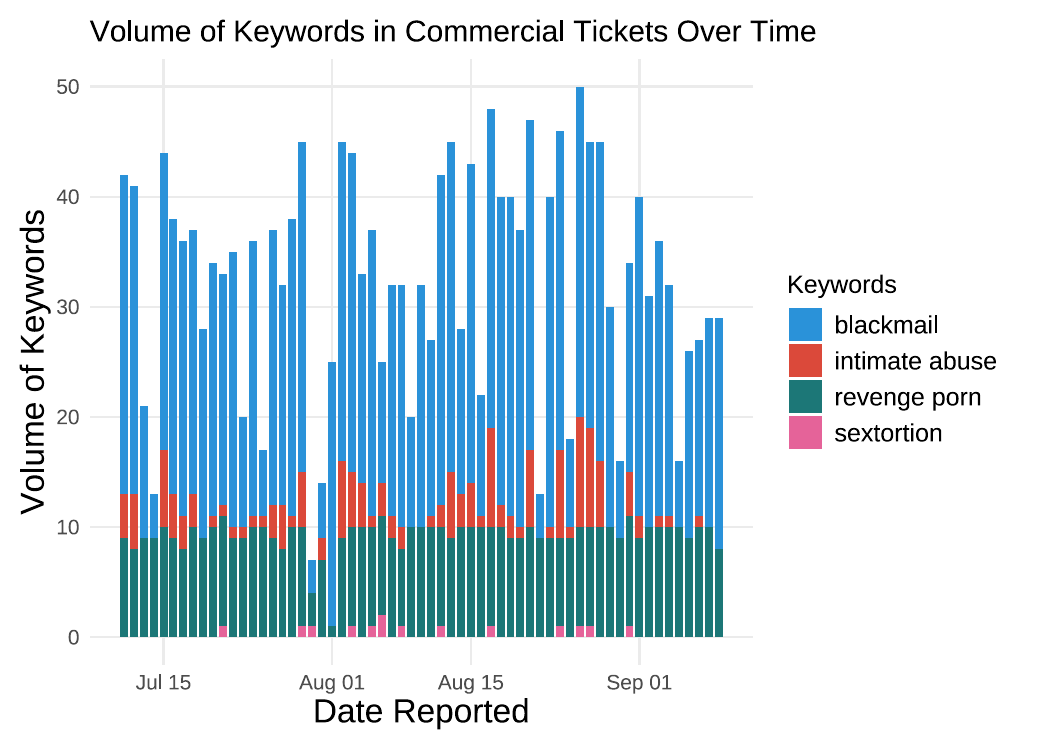}
    \caption{Keywords used in commercial tickets.}
    \label{fig:rq1-4}
\end{figure}

\begin{figure}[H]
    \centering
    \includegraphics[width=0.7\linewidth, trim={ 0 40 0 50}, clip]{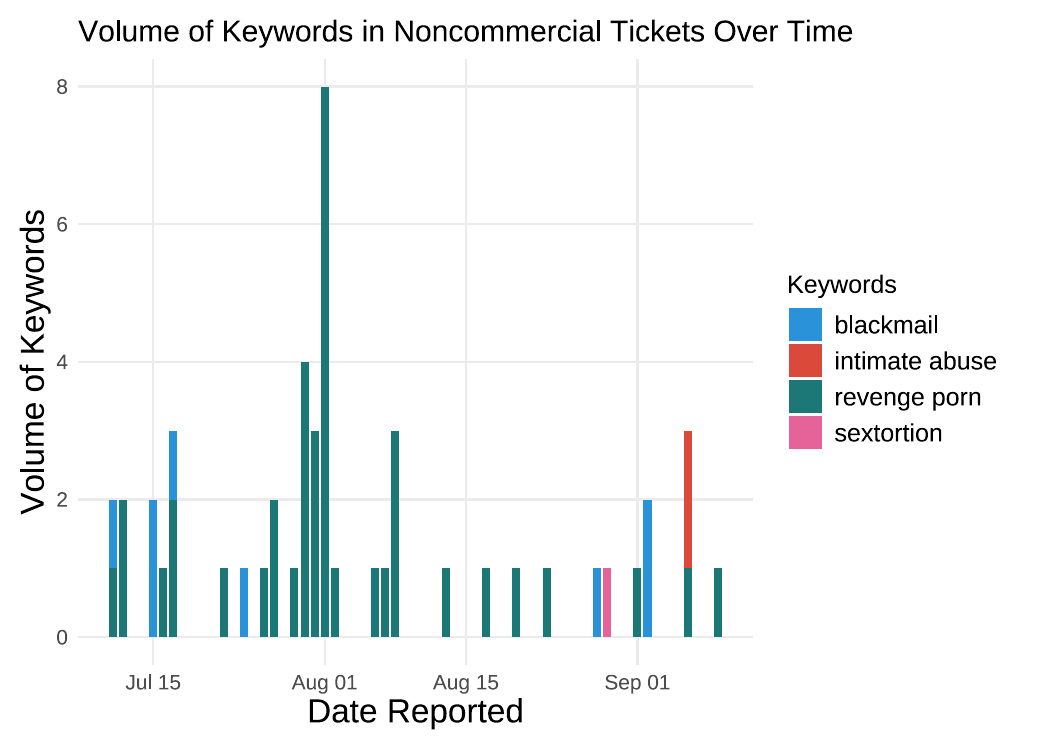}
    \caption{Keywords used in non-commercial tickets.}
    \label{fig:rq1-5}
\end{figure}

\keywcommercial
\keywnoncommercial
    
\end{appendices}

\end{document}